\documentclass[preprint,12pt]{elsarticle}

\usepackage{amssymb}
\usepackage{amsmath}

\usepackage{graphicx}
\usepackage{longtable}
\usepackage{caption}
\usepackage{upgreek}
\usepackage{xcolor}
\usepackage[bookmarks=false]{hyperref}
\hypersetup{%
  colorlinks=true,
  linkbordercolor=red,
  pdfborderstyle={/S/U/W 1}
}

\journal{Physics Reports}

\begin{document}

\begin{frontmatter}



\title{Advancing time- and angle-resolved photoemission spectroscopy: The role of ultrafast laser development}


\author{MengXing Na $^{a,b}$}

\affiliation{organization={Quantum Matter Institute, University of British Columbia},
            city={Vancouver},
            state={BC},
            country={Canada},
            postcode={~V6T 1Z4}
            }

\author{Arthur K. Mills $^{a,b}$}
\author{David J. Jones $^{a,b}$}

\affiliation{organization={Department of Physics and Astronomy, University of British Columbia},
            city={Vancouver},
            state={BC},
            country={Canada},
            postcode={~V6T 1Z1}
            } 

\begin{abstract}
In the last decade, there has been a proliferation of laser sources for time- and angle-resolved photoemission spectroscopy (TR-ARPES), building on the proven capability of this technique to tackle important scientific questions. In this review, we aim to identify the key motivations and technologies that spurred the development of various laser sources, from frequency up-conversion in nonlinear crystals to high-harmonic generation in gases. We begin with a historical overview of the field in Sec.\,\ref{Sec: Introduction}, framed by advancements in light source and electron spectrometer technology. An introduction to the fundamental aspects of the photoemission process and the observables that can be studied is given in Sec.\,\ref{Sec: Fundamental}, along with its dependencies on the pump and probe pulse parameters. The technical aspects of TR-ARPES are discussed in Sec.\,\ref{Sec: Technical}. Here, experimental limitations such as space charge and resultant trade-offs in source parameters are discussed. Details of various systems and their approach to these trade-offs are given in Sec.\,\ref{Sec: Laser survey}. Within this discussion, we present a survey of TR-ARPES laser sources; a meta-analysis of these source parameters showcases the advancements and trends in modern systems. Lastly, we conclude with a brief discussion of future directions for TR-ARPES and its capabilities in elucidating equilibrium and non-equilibrium observables, as well as its integration with micro-ARPES and spin-resolved ARPES (Sec.\,\ref{Sec: outlook}).
\end{abstract}

\begin{graphicalabstract}
\includegraphics[width=1\columnwidth]{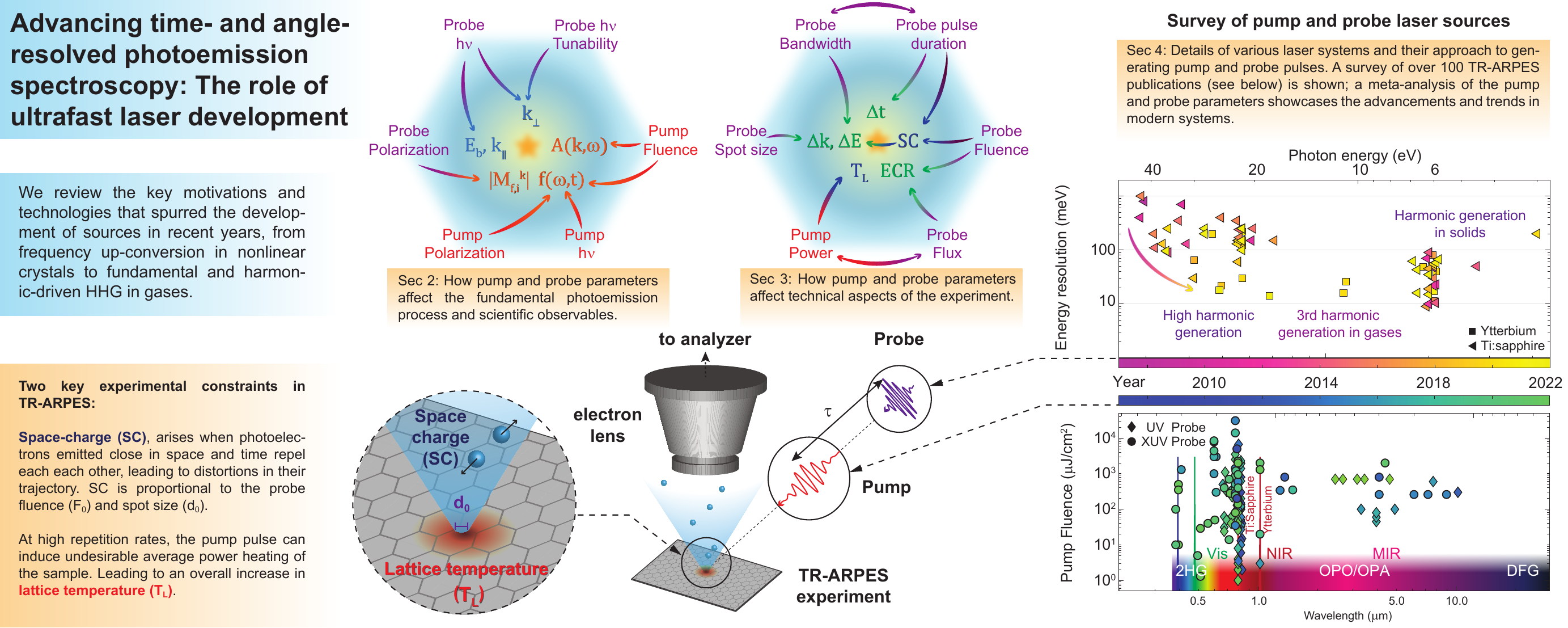}
\end{graphicalabstract}


\end{frontmatter}
\tableofcontents
\pagebreak
\section{Introduction}
\label{Sec: Introduction}
Angle-resolved photoemission spectroscopy, or ARPES, is a photon-in, electron-out experimental technique that has become a cornerstone in modern studies of condensed matter physics \cite{Damascelli2004}. Its ability to access the electronic band structure and many-body interactions makes ARPES an unparalleled probe of crystalline materials. By combining ARPES with pump-probe spectroscopy, these investigations are extended into the time domain. Access to the dynamics of the momentum-resolved energy landscape adds important capabilities in condensed matter research. The optical excitation of electrons into unoccupied states allows for the mapping of the electronic structure above the Fermi energy, which ARPES cannot access; such information can provide important experimental benchmarks for first-principles calculations \cite{Hao2010, Sobota2013, Sobota2014, Yang2017b, Puppin2022}. Interactions that have different timescales, such as electron-electron and electron-phonon scattering, can be discerned through the relaxation process \cite{Allen1987, Lisowski2005, Perfetti2007, Kirchmann2010}, which can offer important clues to the underlying interactions behind equilibrium electronic phases \cite{Schmitt2008, Rohwer2011, Hellmann2012, DalConte2012}. Moreover, optical pulses are versatile tools with which one can control the electronic properties of the material \cite{Armitage2014}. For instance, ultrafast carrier redistribution has been demonstrated using two-color current injection \cite{Bas2015}, surface photo-voltage \cite{Yang2014, Neupane2015, Michiardi2022}, and single-cycle THz pulses \cite{Reimann2018}. Photo-induced phase transitions have provided insights into the nature of unconventional superconductivity \cite{Perfetti2006, Smallwood2012, Giannetti2016, Boschini2018}, charge-order \cite{Schmitt2008, Rohwer2011}, metal-to-insulator transitions \cite{Hellmann2012}, and excitonic insulators \cite{Monney2016, Mor2017, Hedayat2019, Tang2020}, among others. Lastly, optically-induced non-equilibrium states --from excitons \cite{Weinelt2004, Madeo2020, Mori2023} to Floquet-Bloch states \cite{Wang2013, Mahmood2016, Zhou2023, Ito2023}-- have been experimentally observed. 

The success of ARPES is enabled by technological advancement in two key areas: the development of electron spectrometers with high energy and momentum resolution and the development of light sources, most prominently third-generation synchrotrons \cite{Damascelli2003}. From this foundation, advances in time-resolved ARPES (TR-ARPES) have followed from the development of high-intensity ultrafast lasers and the ability to produce synchronized femtosecond pump and probe pulses, with the probe photon energy exceeding the work function of the material (typically 4 to 5~eV). TR-ARPES experiments using such femtosecond laser sources are preceded by two notable and related photoemission techniques employing photon energies less than the work function. Firstly, thermally-assisted photoemission --also known as multi-photon thermionic photoemission-- uses intense femtosecond pulses to drive the electron bath to temperatures exceeding 10$^3$~K, resulting in photoemission from low-order processes \cite{Fujimoto1984, Weidele1995}. Secondly, the femtosecond pulses with higher photon energy (2-5~eV) were used to drive multi-photon photoemission events \cite{Giesen1985, Schoenlein1988, Girardeau1995, Aeschlimann1995, Damascelli1996, Petek1997, Zhu2004, Fauster2007}. These early experiments demonstrated a pump-induced thermal non-equilibrium between the electrons and the lattice and are foundational for a variety of experiments today. In this review, we focus primarily on the development of ultrafast laser sources for TR-ARPES; the interdependency of laser source parameters and new developments in electron spectrometers are highlighted when it is relevant. In some cases, developments in laser and analyzer technology occur together; at other times, technology is adopted from elsewhere such as equilibrium ARPES. Modern TR-ARPES is also preceded by TR-PES systems without angular resolution. For the purposes of this review, we will define ``angle-resolved" systems as those with parallel detection of energy and angle in which dispersive features are observed. Further, we constrain our discussion of TR-ARPES to those pump-probe sources in which the probe (pump) photon energy is above (below) the work function of the material, that is, where the photoemission matrix element only concerns the interaction between the probe pulse and the electron.

\subsection{Historical Development}
\label{Sec: History}
\begin{figure}[t!]
    \centering
    \includegraphics[width=0.97\columnwidth]{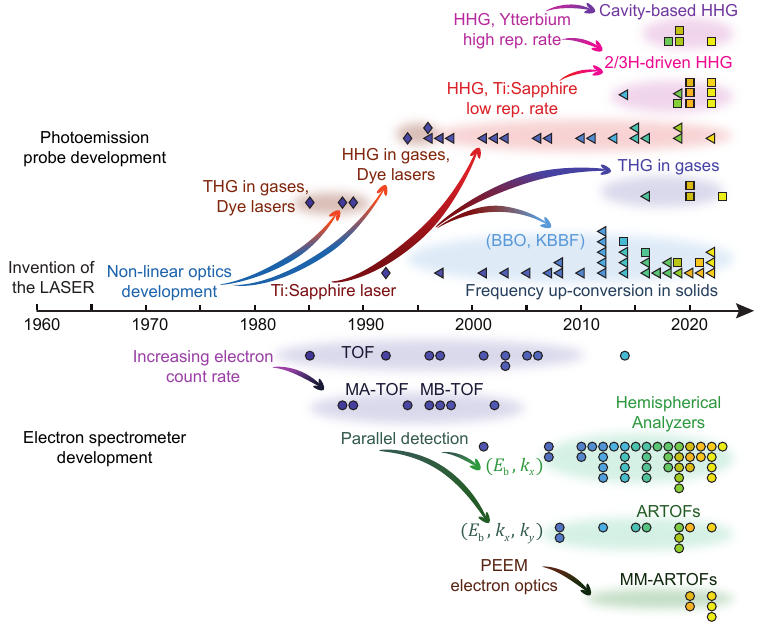}
    \caption{\textbf{A sketch of the historical development of time- and angle-resolved photoemission spectroscopy (TR-ARPES).} The probe source (diamonds, triangles, and squares denote the dye, Ti:sapphire, and ytterbium-doped  laser platforms) and its corresponding electron spectrometer (circles) are plotted on the timeline by publication year. Each pair of points corresponds to a distinct TR-ARPES setup and the references are tabled in Tab.\,\ref{Tab: history}. Although the TR-ARPES probe had its roots at higher photon energies in the 1980s, the proliferation of TR-ARPES  around 2005 was prompted by the development of Ti:sapphire-based 6-eV sources and parallel detection schemes from hemispherical analyzers. Subsequent investment into higher photon energies, repetition rates, and energy resolutions was made on the probe side through harmonic-driven and cavity-based HHG. On the analyzer side, TOFs were initially employed for their higher electron throughput since repetition rate and electron count rates were low. Additional parallelization developed in the form of angle-resolved time-of-flight spectrometers (ARTOFs). Momentum-microscope (MM)-ARTOFs give additional spatial resolution. T/FHG: Third/fourth harmonic generation. HHG: high-harmonic generation. TOF: Time-of-flight. MA: Multi-anode. MB: Magnetic-bottle. PEEM: Photoemission electron microscopy. MM: Momentum-microscope.}
    \label{Fig: history}
\end{figure}
The historical evolution of TR-ARPES technology naturally reflects the collective advances in ultrafast lasers, nonlinear optics, and the development of electron spectrometers, as sketched in Fig.\,\ref{Fig: history}. The first demonstration of laser-based photoemission from pump-excited states used Nd:YAG (neodyminum-doped yttrium aluminum garnet) pumped dye lasers. 

In the late 1980s, photoemission was performed with picosecond-duration 10.7~eV pulses generated via UV-driven third harmonic generation (THG) in gases \cite{Haight1985, Haight1988}. Probe pulses were produced using frequency upconversion in nonlinear crystals such as BBO in the early 1990s \cite{Fann1992}. Subsequently, high-order harmonic generation (HHG) was demonstrated in the mid-1990s with two different sources with subpicosecond (800~fs and 150~fs) pulses \cite{Haight1994}. In these TR-PES experiments, the pulses were amplified through multiple stages of dye cells to produce 0.3-0.5~mJ pulse energy at 550-720~nm \cite{Haight1988}. At a repetition rate of 100-200~Hz, photoelectron flux is low and constrained by space charge effects (detailed discussion in Sec.\,\ref{Sec: space charge}). Hence, the time-of-flight (TOF) spectrometer was chosen to capture all electrons in a single laser shot. The development of a multi-anode detector further increased sensitivity and allowed some angular information to be captured. Refinements of the laser technology later produced pulses with 0.5~mJ energy and 800~fs, which enabled high-order harmonic generation up to 25~eV. 

Using a different laser system, intracavity-prism dispersion compensation in the dye laser reduced the pulse duration to 150~fs, enabling efficient HHG at higher photon energies (up to 80~eV) and increased electron counts at a repetition rate of 540~Hz  \cite{Haight1994, Haight1996}. While the thermionic and multi-photon photoemission experiments were used to study image-potential states in metals, these first TR-PES systems unlocked intervalley scattering dynamics in semiconductors \cite{Rettenberger1996, Rettenberger1997}. Refinements of these first systems foreshadow the development of TR-ARPES for the next three decades, such as increasing the electron count rate through parallel detection and higher repetition rate.

In the 1990s, the advent of femtosecond (fs) titanium-doped sapphire (Ti:sapphire) laser systems led to the wide adoption of ultrafast laser systems and their use in a multitude of experimental studies, including photoemission spectroscopy \cite{Karlsson1996}. The intense, short pulses --enabled by the large gain bandwidth of Ti:sapphire-- were particularly suitable for pushing HHG to its extremely short pulse limits. The photoemission signal from laser-dressed helium atoms \cite{Glover1996} (equivalent to laser-assisted photoemission or Volkov states) demonstrated cross-correlations of pump and probe in the range of 100 to 200~fs \cite{Bouhal1997, Bouhal1998}. These experiments were accompanied by a magnetic-bottle TOF detector, which -- though not angle-resolved -- captures the entire solid angle of electrons and further increases the electron count rate \cite{Kruit1983}.

The first Ti:sapphire amplifier systems operated at low repetition rates ($\approx$~10~Hz) to leverage pulsed Nd:YAG pump sources. Later, Ti:sapphire regenerative amplifiers pushed the repetition rates into the low kHz regime \cite{Rudd1993, Backus1995}. Studies on solids and gases followed, using magnetic-bottle and regular TOF detectors \cite{Siffalovic2001, Nugent-Glandorf2002, Bauer2003, Miaja-Avila2006}. While efficient, these detectors lacked angular resolution and the ability to resolve the electronic dispersion --a crucial feature unique to photoemission as a probe of crystalline solids. The integration of two-dimensional hemispherical analyzers (HA) enabled parallel detection of the photoelectron energy and emission angle \cite{Valla1999}. Developed for static experiments with continuous sources, HAs were not initially used with kilohertz femtosecond sources, given the space charge limitations at low repetition rates. The integration of the HA with time-resolved two-photon photoemission showed the parabolic dispersion of an image potential state in 2005 \cite{Rohleder2005}, and the integration of HA with a 1 kHz HHG source demonstrated the ability to capture momentum-resolved features in 2007 \cite{Mathias2007}. Shielding and electron optics were subsequently adapted for lower kinetic energy electrons \cite{Koralek2007}, enabling the employment of 6-eV sources discussed below.

Based on the fourth harmonic generation (FHG) in nonlinear optical crystals, most prominently beta-barium borate (BBO), the uptake of 6-eV sources was fast and they quickly outnumbered HHG sources. The high peak intensity of Ti:sapphire laser systems is well suited to drive efficient harmonic generation in nonlinear crystals, and --as 6-eV sources are in the UV-- they do not require vacuum or gases to implement. Although some 6-eV sources have been demonstrated with TOFs previously \cite{Perfetti2007, Schmitt2008, Link2001, Rhie2003, Lisowski2004}, their proliferation was largely in the context of HAs due to the added ability to study the evolution of electron dispersion \cite{Graf2011, Faure2012, Sobota2012, Wang2012, Crepaldi2012, smallwood2012RSI, Ishida2014, Wegkamp2014, Andres2015}. Straightforward to build and maintain, 6-eV sources have been critical in establishing TR-ARPES as a pillar in the study of the dynamic electronic structure of solids and remain a productive fixture in TR-ARPES labs. The key limitation of 6-eV sources is that the low photon energy limits the $\boldsymbol{k}_\parallel$ accessible by photoemission (details in Sec.\,\ref{Sec: Fundamental}). Recently, potassium beryllium fluoroborate (KBBF) has pushed the limit up to 7~eV \cite{Bao2022, Zhong2022}, but to increase the photon energy further; one must move past the transparency window of solids. Sources based on UV-driven THG in gases have also been developed \cite{Cilento2016, Zhao2017, Lee2020, Peli2020, Kawaguchi2023}, albeit less abundantly than 6~eV. Although the modest increase in photon energy (up to 11~eV) is relevant for many materials, it is insufficient to span the Brillouin Zone (BZ) of materials such as graphene. The technical complexity accompanying the use of gases may also be a barrier. Lastly, THG seeded by the second harmonic of Ti:sapphire sources is inefficient due to phase-matching conditions at the relevant wavelengths. With the maturity of commercial Yb-doped laser platforms, we may see the number of probe sources based on THG in gases increase in the near future.

To achieve full coverage of the BZ of all materials, one must return to HHG sources. As discussed previously, the HHG process produces higher photon energy and is typically easier to implement when driven with short pulses and high peak electric fields. Many early sources leaned on these aspects and employed driving pulses compressed down as far as 10~fs \cite{Petersen2011, Frassetto2011, Frietsch2013, Eich2014, Rohde2016}. While these sources were scientifically productive, the large corresponding bandwidth and poor energy resolution were cumbersome for detailed studies of the electron dispersion and spectral function. To access spectral features at lower energy scales, a push towards higher energy resolution was made. Notably, the HHG process driven by the second or third harmonic of a Ti:sapphire is much more efficient than that driven by the fundamental, permitting HHG sources to operate at higher repetition rates (details in Sec.\,\ref{Sec: Laser survey}). Energy resolutions of 30 to 60~meV, on par with those achieved by 6-eV sources, have been demonstrated \cite{Sie2019, Buss2019}. The larger spectral separation between harmonics (relative to fundamental-driven HHG) also facilitates efficient spectral separation of ultrashort pulses with multilayer mirrors or time-preserving monochromators, which employ off-plane low-density gratings \cite{Frassetto2011}. Due to the high pulse energy requirements, Ti:sapphire HHG sources have also struggled to increase in repetition rate, often operating on the order of 1 to 10~kHz in contrast to the 100~kHz routinely achieved by 6-eV sources. Recently, high-power ytterbium (Yb)-doped fiber and Yb-doped crystal lasers have accelerated the adoption of high repetition rate sources up to 100~MHz. Yb-doped fiber lasers were first used in 6-eV generation at a repetition rate of 95~MHz \cite{Ishida2016}. Yb-doped fiber and Yb-doped crystal lasers have also been used to seed UV-driven THG systems \cite{Lee2020,Peli2020}, and HHG systems in a variety of configurations at 100~kHz \cite{Puppin2019, Cucini2020, Guo2022}, 1~MHz \cite{Madeo2020, Keunecke2020} and $>$10~MHz \cite{Mills2019, Corder2018} and have recently demonstrated excellent energy resolutions (14~meV \cite{Guo2022}).

Despite increasing repetition rates, the electron count rate still left much to be desired, particularly in time-resolved experiments where photoemission intensity above the Fermi level is low. Compounding this issue, the collection efficiency of the HA hindered by the entrance slit, as it measures only one slice of the three-dimensional photoelectron distribution per laser shot. The next development was the added angular resolution to the TOF detector (ARTOF), which allows the full in-plane electron dispersion to be captured in parallel \cite{Ovsyannikov2013}. The proliferation of ARTOFs is recent, paired with both 6-eV \cite{Wang2012, Rettig2016, Freutel2019} and HHG sources \cite{Sie2019, Saule2019, Guo2022}, though the latter is more popular for its large momentum space coverage. Lastly, in the 2000s, significant effort was made to integrate the photoelectron momentum microscope lens element with the electron spectrometer \cite{Kotsugi2003, Kromker2008, Chernov2015}. This integration has enabled experiments on the micrometer scale without focusing the probe to micrometer spot size \cite{Keunecke2020, Kutnyakhov2020, Schonhense2021, Heber2022}. These systems (MM-ARTOFs) have been successfully applied to the study of excitons and electron dynamics in exfoliated monolayers and heterostructures of transition-metal dichalcogenides \cite{Madeo2020, Dong2021, Man2021, Wallauer2021, Schmitt2022, Kunin2023} and molecular thin films \cite{Wallauer2021b, Baumgartner2022, Bennecke2023}.

\subsection{Current State of TR-ARPES}
Today, TR-ARPES is a staple in the study of ultrafast electron dynamics. For a review of the scientific output, we refer the reader to the existing scientific reviews \cite{Smallwood2016, Zonno2021, Zhang2022, Boschini2023}. As experimental techniques improve, so does the complexity of the acquired data and the accompanying theoretical descriptions. TR-ARPES experiments are supported and spurred on by a large body of theoretical work which continues to evolve as the sources unlock new excitation and measurement regimes \cite{Allen1987, Freericks2009, Sentef2013, Braun2015, Kemper2017, Kemper2018, Kim2020, Freericks2021, Eckstein2021, Schuler2021, Caruso2022, Degiovannini2022}.

While we have highlighted efforts to improve the probe pulse in this section, recent efforts have also been made to increase the flexibility of the pump pulse to access different excitation regimes and observables. With the proliferation of TR-ARPES sources, this technique stands ready to tackle many challenging questions at the forefront of condensed matter physics. To underscore its potential, one needs to look no further than the host of theoretical works outlining the use of TR-ARPES in the study of electron-boson coupling \cite{Sentef2013, DeGiovannini2020}, bright and dark excitons \cite{Mori2023, Rustagi2018, Christiansen2019, Sangalli2021}, and excitonic insulators \cite{Weinelt2004, Perfetto2020}, to name but a few. The light-induced anomalous Hall effect \cite{Sentef2015, McIver2020},  Floquet physics \cite{Sentef2015, Farrell2016, Schuler2020}, coherences between states \cite{Rustagi2019, Marini2022}, light-induced superconductivity \cite{Kennes2017}, and fluctuation of order parameters \cite{Schwarz2020} are a few more experiments that have been proposed. Although other ultrafast spectroscopic techniques have been applied to these questions, a direct probe of the momentum-resolved electronic structure in its non-equilibrium phase is still highly relevant.

With this in mind, Sec.\,\ref{Sec: Fundamental} of this manuscript reviews the fundamental aspects of the photoemission process and how the probe source parameters factor into the extraction of the electronic dispersion and the interpretation of the ARPES intensity. Then in Sec.\,\ref{Sec: Technical}, we discuss three primary technical trade-offs in which light-source considerations are critical: time-energy resolution, space charge distortions, and thermal dissipation. Together, these provide an overview for those experts considering the design of a laser source for photoemission experiments. In Sec.\,\ref{Sec: Laser survey}, we shift our focus to a survey of laboratory-based laser sources. This discussion is supported by a survey of over 100 TR-ARPES studies from which some interesting trends emerge. We continue with a discussion of technical details and considerations behind the laser systems currently being used to produce the pump and probe pulses for TR-ARPES experiments. This section provides an overview of existing sources and capabilities for those experts seeking to design specific experiments. We conclude with a discussion of some of the emerging directions in TR-ARPES in Sec.\,\ref{Sec: outlook}, such as capabilities enabled by advances in source technology and integration of ultrafast light sources with micro- and spin-resolved ARPES and the unique opportunities offered by free-electron lasers.

\section{Fundamental aspects of ARPES}
\label{Sec: Fundamental}

\begin{figure}[ht!]
  \begin{center}
    \includegraphics{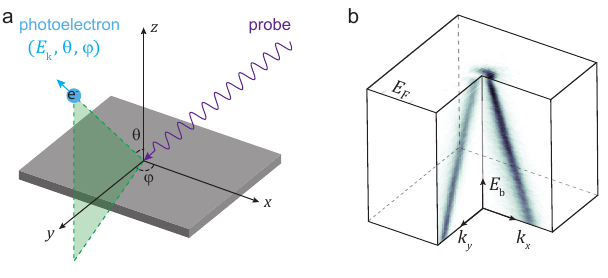}
    \caption[Energy and momentum conservation in angle-resolved photoemission spectroscopy]{\textbf{Energy and momentum conservation in angle-resolved photoemission spectroscopy.} \textbf{(a)} ARPES experimental geometry. An incoming photon (probe) ejects an electron from the sample with emission angles $\theta$ and $\phi$ and kinetic energy $E_\text{k}$. Since the energy and in-plane ($x-y$) momentum is conserved during the photoemission process, one can map $(E_\text{k}, \theta, \phi)$ onto $(E_\text{b}, \boldsymbol{k}_\parallel)$. \textbf{(b)} ARPES spectra measured on hydrogen-intercalated graphene. The measured $(E_\text{k}, \theta, \phi)$ transform into $(E_\text{b}, k_x, k_y)$ via conservation equations.}
   \label{Fig: ARPES_geometry} 
  \end{center}
\end{figure}

Many review papers and textbooks on ARPES have been written throughout the years, and we refer the reader to these past works \cite{Damascelli2004, Damascelli2003, Zhang2022, Cardona1978, Kevan1992, Hufner2013, Yang2018, Suga2021, Sobota2021}. In this section, we discuss how various parameters of the probe relate to the photoemission process. Specifically, we consider the energy-momentum conservation equations and the measured ARPES intensity.

\subsection{Conservation equations}
In photoemission, a high-energy photon excites the material, inducing the photoemission of an electron \citep{Damascelli2004}. The energy conservation of the photoemission process is
\begin{equation}
\begin{split}
    E_\text{k}&=h\nu-\mathit{\Phi}-E_\text{b},
    \label{Eq: Ek}
\end{split}
\end{equation}
where $E_\text{k}$ is the kinetic energy of the photoelectron, $h\nu$ is the energy of the photon, $\mathit{\Phi}$ is the work function of the material, and $E_\text{b}$ is the binding energy of the electron relative to the Fermi-energy of the material. Due to translation symmetry in the plane of the sample surface ($x$-$y$ plane in Fig.\,\ref{Fig: ARPES_geometry}a), in-plane momentum ($\boldsymbol{k}_\parallel$) is conserved during the photoemission process. In the sample reference frame, the conservation equation is
\begin{equation}
    \begin{split}
        \boldsymbol{k}_\parallel=\sqrt{\frac{2mE_\text{k}}{\hbar^2}}\sin(\theta),\\
    \end{split}
    \label{Eq: k_parallel}
\end{equation}
where $\theta$ is the polar angle of the emitted photoelectron. The azimuth angle $\phi$ then decomposes $\boldsymbol{k}_\parallel$ into $k_x$ and $k_y$ components. This angle-to-$\boldsymbol{k}_\parallel$ conversion in the sample reference frame is shown for its clarity; however, in the lab frame, the sample orientation is often manipulated with respect to a stationary detector. Therefore, the angle-to-$\boldsymbol{k}_\parallel$ conversion depends on the exact experimental setup and the modes of operation of the electron analyzer \cite{Ishida2018}. An example of ARPES spectra -- measured on hydrogen-intercalated graphene-- is shown in Fig.\,\ref{Fig: ARPES_geometry}b. The characteristic linear dispersion at the $K$ point is observed. The small hole pocket (due to the slight p-doping given by the hydrogen intercalation) can be seen on the Fermi surface (the constant energy contour at the Fermi energy $E_\text{F}$).

\begin{figure}[t!]
  \begin{center}
    \includegraphics{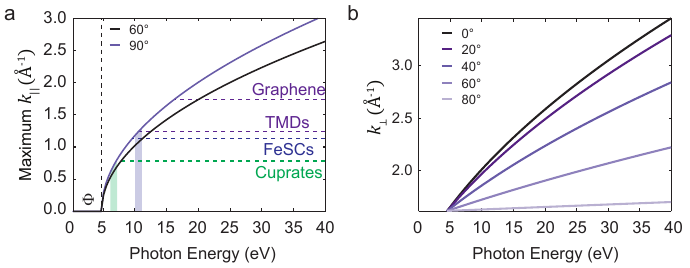}
    \caption[Momentum space coverage is given by various photon energies]{\textbf{Momentum space coverage given by various photon energies.} \textbf{(a)} The maximum accessible parallel momenta $\boldsymbol{k}_\parallel$ as a function of photon energy for emission angles $60^{\circ}$, and $90^{\circ}$. Horizontal dashed lines indicate the BZ boundary of several classes of materials. The available photon energies from 6-eV sources and THG in gases are indicated by shaded regions green and blue, respectively. The work function is $\mathit{\Phi}=4.5$~eV. TMD: Transition-metal dichalcogenides. FeSCs: Iron-based superconductors. \textbf{(b)} The out-of-plane crystal momentum ($k_\perp$)as calculated from Eq.\,\ref{Eq: k_perp}, using $V_0=10$~eV and $\mathit{\Phi}=4.5$~eV.}
  \label{Fig: Photon energy}  
  \end{center}
\end{figure}

Eq.\,\ref{Eq: k_parallel} sets a limit on the $\boldsymbol{k}_\parallel$ accessible by photoemission. The emission angle $\theta=0^{\circ}$ corresponds to normal emission ($\boldsymbol{k}_\parallel=0$). ARPES setups are conservatively capable of collecting photoelectrons with $\theta\leq 60^{\circ}$. The integration of the momentum-microscope lens column, with large extraction voltages and short working distances, enables the entire $90~^{\circ}$ to be captured \cite{Karni2023}. In Fig.\,\ref{Fig: Photon energy}a, we plot $\boldsymbol{k}_\parallel^\mathrm{max}$ against the photon energy for $60^{\circ}$, and $90^{\circ}$, assuming a work function of $\mathit{\Phi}=4.5$~eV. The BZ edge for several families of materials is marked (horizontal dashed lines). The widely adopted frequency up-conversion in nonlinear optical crystals such as BBO and KBBF typically produces photons in the range of $6-7$~eV (shaded green region) and are fairly limited in momentum space coverage. THG in gases (shaded blue region) increases the range up to 11~eV, which can reach the BZ edge of TMDs.

Since the out-of-plane translation symmetry is broken at the sample surface, $k_\perp$ or $k_z$ --as defined in Fig.\,\ref{Fig: ARPES_geometry}a-- is not conserved. One approach to determine $k_\perp$ is to use a nearly-free electron model to approximate the final state. Then the out-of-plane crystal momentum can be written as \cite{Damascelli2004}:
\begin{equation}
    k_\perp=\sqrt{2m/\hbar^2(E_\text{k}\cos^2(\theta)+V_0)},
    \label{Eq: k_perp}
\end{equation}
where $V_0$ is commonly referred to as the inner potential. To map the out-of-plane dispersion, ARPES spectra are acquired for a range of photon energies in steps of $\approx 1$~eV that spans multiple BZ in the $k_\perp$ direction (see Fig.\,\ref{Fig: Photon energy}b). The known periodicity of the BZ calculated from the crystal structure is then fitted to a periodic feature of ARPES spectra with Eq.\,\ref{Eq: k_perp} to obtain $V_0$. This method works well at high photon energies ($>$100~eV) where the electron mean-free-path is large, and the final state is well approximated by a nearly-free electron state. However, it is also often used at lower photon energies, where one must be more careful in interpreting the results. 

Ultimately, Eq.\,\ref{Eq: k_perp} concerns the tunability of the light source. In static ARPES experiments, the photon energy dependence is typically obtained at a synchrotron \cite{Damascelli2004}. TR-ARPES is not generally used to map the out-of-plane dispersion, but the photon energy tunability may be relevant with respect to accessing regions of $k_\perp$. In this respect, the 6-eV sources provide continuous tunability over a limited range given by the phase-matching conditions of the nonlinear crystal (5.5-6.7~eV for BBO, up to 7~eV for KBBF) \cite{Zhong2022, Bao2022}. On the other hand, HHG sources provide a larger range of photon energies but without continuous tunability. In HHG, neighbouring harmonics are separated by twice the energy of the fundamental driving field. For sources driven with Yb-doped or Ti:sapphire-based gain media, the spectral separation is approximately 2.4~eV and 3.1~eV, respectively, which may provide some utility for accessing the out-of-plane dispersion. The spectral separation for HHG driven by the 2$^\text{nd}$ or 3$^\text{rd}$ harmonic of the fundamental is too large to be useful in this respect.

\begin{figure}[t!]
  \begin{center}
    \includegraphics{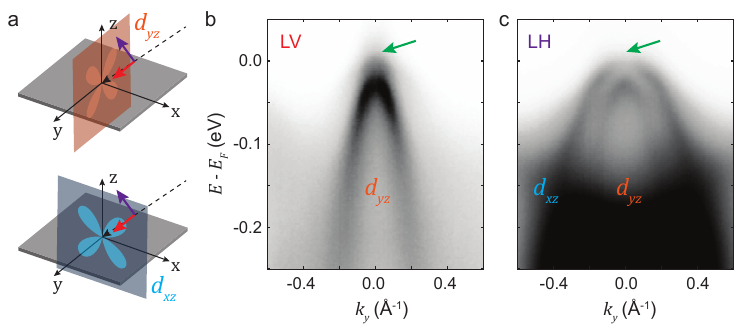}
    \caption[Effects of light polarization and photoemission matrix elements on ARPES spectra of FeSe]{\textbf{Effects of light polarization and photoemission matrix elements on ARPES spectra of FeSe.} Figure adapted from Day \textit{et al.} \cite{Day2019}.  \textbf{(a)} The experimental geometry: The linear vertical (LV) light is polarized along the $y$ and selectively probes the $d_{yz}$ orbital (top). The linear horizontal (LH) light has components along $x$ and $z$ and can probe both $d_{xz}$ and $d_{yz}$ characters. \textbf{(b)} ARPES intensity from the valence band of FeSe acquired with $h\nu=37$~eV and  120~K. The orientation is along the $\Gamma-$M direction. The polarization is set to LV for \textbf{(b)} and LH for \textbf{(c)}, allowing for photoemission from states of different orbital characters. Green arrows indicate a spectral weight transfer from LH to LV spectra as spin-orbit coupling mixes the orbital character.}
  \label{Fig: Matrix elements}  
  \end{center}
\end{figure}

\subsection{Intensity}
\label{Sec: ARPES intensity}
The total ARPES intensity is given by
\begin{equation}
    \begin{split}        I(\boldsymbol{k},\omega)&\propto\left|M_{f,i}^{\boldsymbol{k}}\right|^2A(\boldsymbol{k},\omega)f(\omega),
    \end{split}
    \label{Eq: ARPES intensity2}
\end{equation}
where $f(\omega)$ is the occupation function at electronic temperature $T_e$, such that ARPES only probes occupied states, $M_{f, i}^{\boldsymbol{k}}$ is the photoemission matrix element, and $A(\boldsymbol{k},\omega)$ is the one-electron removal spectral function. The photoemission matrix element accounts for the polarization of the photon as well as the symmetry of the initial state ($\phi_i^{\boldsymbol{k}}$) and final state ($\phi_f^{\boldsymbol{k}}$), and is calculated as $|M_{f, i}^{\boldsymbol{k}}|^2\propto |\langle\phi_f^{\boldsymbol{k}}|\epsilon\cdot\mathbf{x}|\phi_i^{\boldsymbol{k}}\rangle|^2$, where $\epsilon$ is the polarization unit vector \cite{Damascelli2004}. Understanding of the photoemission matrix element has advanced in recent years, such that simulations can capture much of the matrix element effects in the photoemission intensity \cite{Day2019, Moser2017}. This factor can strongly modulate the photoemission intensity, even entirely suppressing intensity from certain bands, such as the well-known example of the ``dark corridor" in graphene \citep{Gierz2011}. 

The effect of photoemission matrix elements on FeSe is demonstrated in Fig.\,\ref{Fig: Matrix elements}, adapted from Day \textit{et al.} \cite{Day2021}. The experimental geometry is shown in Fig.\,\ref{Fig: Matrix elements}a, the linear vertical (LV) polarization is in-plane and along $\hat{y}$, while the linear horizontal (LH) polarization is out-of-plane, with components along $\hat{x}$ and $\hat{z}$. In Fig.\,\ref{Fig: Matrix elements}b and c, we see that linear vertical (LV) polarization selectively photoemits the band with $d_{yz}$ orbital character, while linear horizontal (LH) polarization can probe both  $d_{yz}$ and $d_{xz}$ bands. Interestingly, at the $\Gamma$ point, spin-orbit coupling mixes the character of the $d_{xz}$ and $d_{yz}$ bands, which results in a transfer of spectral weight from the LH to the LV spectra (green arrows). 

The polarization of light is an additional tuning knob to study the electron wavefunction \cite{Zhu2013, Cao2013, Beaulieu2020, Beaulieu2021b, Schuler2022}. The symmetry sensitivity of light-polarization can be used in dichroism experiments to study quantities such as the orbital angular momentum (OAM) \cite{Park2012, Park2012b, Sunko2017}, the non-Abelian Berry curvature \cite{Cho2018, Schuler2020b, Cho2021}, and for detecting transient topological phases \cite{Schuler2020}. By combining circularly polarized light with a spin-polarized detector, one can combine orbital and spin selectivity to study spin-orbit coupling throughout the BZ \cite{Zhu2013, Veenstra2013, Day2018, Zwartsenberg2022, DiSante2023}. However, it is prudent to mention that these dichroism experiments are difficult to interpret. Photoemission matrix elements are sensitive to not only the initial state, but also the final state, where the free-electron approximation may fail \cite{Mulazzi2009, Scholz2013, Crepaldi2014}. Furthermore, the surface-sensitivity of ARPES leads to a layer-dependent mixture of various orbitals \cite{Zhu2013}. Therefore, while the polarization tunability of the probe in TR-ARPES is an important consideration, the study of dichroism in ARPES and TR-ARPES requires great care and support from theoretical simulations of the one-step model \cite{Dauth2016, Kruger2022}. To date, there have been relatively few experimental results in TR-ARPES that leverage probe polarization. Technical challenges to this implementation are discussed in Sec.\,\ref{Sec: ExpDrivenParams}.

The electronic band structure and interactions are encoded in the one-electron removal spectral function $A(\boldsymbol{k},\omega)$, which is a positive definite quantity defined by the imaginary part of the retarded Green's function
\begin{equation}
    A(\boldsymbol{k},\omega)=\frac{-1}{\pi}\frac{\Sigma''_{\boldsymbol{k}}(\omega)}{[\hbar\omega-\epsilon_{\boldsymbol{k}}-\Sigma'_{\boldsymbol{k}}(\omega)]^2+[\Sigma''_{\boldsymbol{k}}(\omega)]^2}.
\end{equation}
\begin{figure}[t!]
  \begin{center}
    \includegraphics{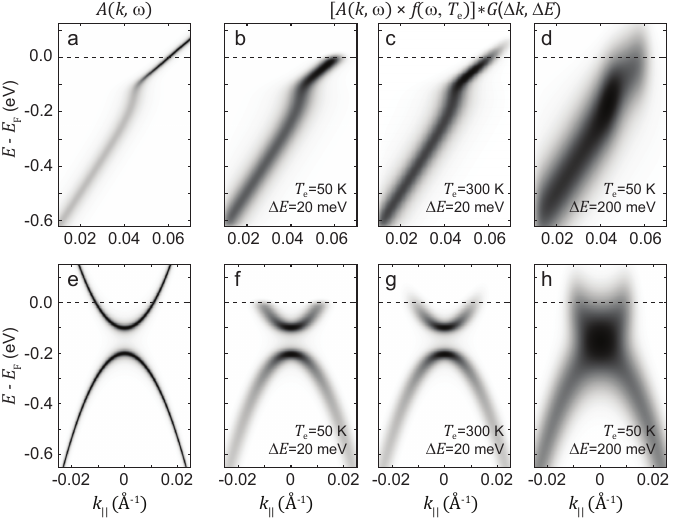}
    \caption[Effects of the energy resolution and thermal broadening on spectral features]{\textbf{Simulation of the effects of the energy resolution and thermal broadening on spectral features.} \textbf{(a)} The spectral function of a linear band crossing $E_\text{F}$, calculated with a small kink at 0.1~eV with amplitude 70~meV representing an electron-boson coupling \cite{Hengsberger1999}. \textbf{(b)-(d)} The spectral function of (a), modified by the Fermi-Dirac distribution and Gaussian convolution to show temperature and resolution effects. \textbf{(e)} The spectral function of two quadratic bands with an energy gap of 100~meV. A 10~meV lifetime broadening is included for visualization purposes. \textbf{(f)-(h)} The spectral function of (b), modified by the Fermi-Dirac distribution and Gaussian convolution to show temperature and resolution effects.}
  \label{Fig: specfun}  
  \end{center}
\end{figure}
In the non-interacting case, the electronic structure is given by the bare-band dispersion $\epsilon_{\boldsymbol{k}}$. As interactions are perturbatively turned on, the corrections to $\epsilon_{\boldsymbol{k}}$ can be expressed in terms of the electron self-energy $\Sigma_{\boldsymbol{k}}(\omega)=\Sigma'_{\boldsymbol{k}}(\omega)+i\Sigma''_{\boldsymbol{k}}(\omega)$. Access to the electron self-energy makes ARPES unique as a probe of many-body correlations. For instance, electron-phonon coupling gives rise to a mass-enhancement ``kink", the magnitude of which can be used to quantify the interaction strength \cite{Hengsberger1999}. A simulation of the spectral function with such a kink is shown in Fig.\,\ref{Fig: specfun}a. Analysis of the spectral function in ARPES typically involves the line shapes of the energy and momentum distribution curves \cite{Damascelli2003}, which are sensitive to broadening contributed by the energy and momentum resolutions \cite{Calandra2007}. In Fig.\,\ref{Fig: specfun}b-d, the effects of temperature broadening (multiplication) and energy resolution (convolution) are shown, and one observes that the extraction of the electron-boson kink is severely hindered at $\Delta E=200$~meV. Similarly, the energy resolution can obfuscate the observation of energy gaps. The effect of temperature and resolution broadening on a 100~meV gap is shown in Fig.\,\ref{Fig: specfun}e-h.

Early in the development of ARPES, the poor energy and momentum resolutions limited ARPES to studies of broad spectral features and the general band dispersion. However, modern ARPES, with its improved resolutions, can sensitively study fine spectral features such as the evolution of small superconducting gaps across the Fermi surface \cite{Ding1996} and band renormalization by electron-phonon coupling \cite{Lanzara2001, Park2007}. TR-ARPES enables the studies of pump-induced changes in the spectral function. The renormalization of the electron-boson kink \cite{Smallwood2012, Zhang2014, Rameau2016} and the quenching of phase coherence in cuprates \cite{Boschini2018} and charge order \cite{Hellmann2012}  are well-known examples. For these types of studies, the energy resolution of the experiment and the fluence of the pump --which affects the electronic temperature-- are important considerations, as demonstrated in Fig.\,\ref{Fig: specfun}.

\begin{figure}[t!]
  \begin{center}
    \includegraphics{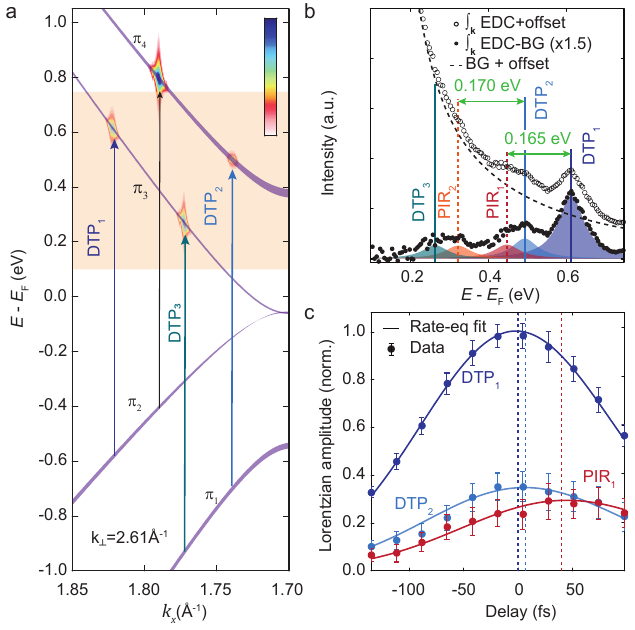}
    \caption[Dynamics of photoexcited carriers in graphite.]{\textbf{Dynamics of photoexcited carriers in graphite.} Figure adapted from Na, Mills \textit{et al.} \cite{Na2019}. \textbf{(a)} The momentum-resolved optical-joint density of states calculated from a tight-binding model showing the available optical transitions in graphite using a pump of 1.19~eV. The energy range in panel b is highlighted in the orange shaded area. \textbf{(b)} The momentum-integrated energy distribution curve ($\int_k$EDC, open circles) is obtained by integrating ARPES intensity along the $\Gamma$-K direction at zero-delay. The background (BG, dashed lines) subtraction highlights a series of peaks (filled circles). The direct optical transition peaks (DTP$_{1, 2, 3}$) are observed alongside secondary peaks (PIR$_{1,2}$) that arise from the emission of strongly-coupled optical phonons ($A_1'$ mode). The phonon energies extracted between the DTP$_1$/PIR$_1$ and DTP$_2$/PIR$_2$ pairs are 0.165~V and 0.170~eV, respectively, as indicated by the green arrow. \textbf{(c)} Evolution of the most prominent peaks. Dark (light) blue correspond to DTP$_1$ (DTP$_2$), and red corresponds to PIR$_1$. The amplitudes are indicative of the population of electrons in each state.}
  \label{Fig: GraphiteSci}  
  \end{center}
\end{figure}
Lastly, we discuss the occupation function $f(\omega)$ in Eq.\,\ref{Eq: ARPES intensity2}. In equilibrium (static) ARPES, $f(\omega)=f_{\text{FD}}(\omega, T_e)$ is a Fermi-Dirac distribution at a finite temperature $T_e$. In TR-ARPES, the pump pulse can induce depletion and population in specific states. The optical-joint density of states (OJDOS) for a particular pump photon energy and polarization has been directly observed in TR-ARPES and shows good agreement with theoretical predictions \cite{Sobota2013, Na2019, Soifer2019}. During photoexcitation, $f(\omega, t)$ is a transient ``non-thermal" distribution. Subsequently, electron-electron, electron-phonon, and electron-magnon scattering processes (among others) redistribute carriers and the absorbed energy \cite{Perfetti2007, Rhie2003, Wang2012, Bovensiepen2007, Cortes2011, Bovensiepen2012, Faure2013, Gierz2013, Johannsen2013, Na2020, Tanimura2021}. 

The evolution of the non-thermal features reflects dominant scattering mechanisms in the system. An example of this is shown in Fig.\,\ref{Fig: GraphiteSci}, adapted from Na, Mills \textit{et al.} \cite{Na2019}. Here, electrons in graphite are shown to preferentially scatter with the strongly-coupled optical phonon ($A_1'$ mode). The optical-joint-density of states illustrates the available optical transitions in Fig.\,\ref{Fig: GraphiteSci}a. In TR-ARPES, the increased occupation of these states manifests as direct-transition peaks (DTP$_{1-3}$) in the momentum-integrated energy distribution curves $\int_k$EDC, obtained by integrating the ARPES spectra in momentum. In graphite, electrons couple strongly to the $A_1'$ optical phonon; the preferential emission of these phonons results in distinct replicas of the DTP in the occupation function, shown in Fig.\,\ref{Fig: GraphiteSci}b. These phonon-induced replicas (PIR$_{1,2}$) are separated from DTP$_{1,2}$ by the energy of the phonon ($\approx 0.165$~eV). Furthermore, the dynamics of PIR are delayed with respect to that of the DTP, as seen in Fig.\,\ref{Fig: GraphiteSci}c. A rate-equation model of these dynamics can be used to extract the characteristic scattering time --and, by extension-- the electron-phonon matrix element. Moreover, multi-temperature models can be used to analyze the evolution of the effective electronic temperature after quasi-thermalization and extract the rate of energy transfer from the electron bath to other degrees of freedom \citep{Allen1987, Perfetti2007, Johannsen2013, Stange2015, Yang2017}. 

In addition to modifying the occupation function, the pump pulse can transiently modify the electronic dispersion and, thus, the spectral function itself. Coherent oscillations of the electronic band structure induced by strong electron-phonon coupling and the melting of equilibrium phases such as charge-density waves and superconductivity are well-known examples \cite{Rohwer2011, Smallwood2012, Gerber2017}. The effect of the pump excitation is strongly material dependent, and pump pulse characteristics, such as photon energy, polarization, and fluence, are important tuning knobs for the study of non-equilibrium physics. A survey of these parameters is discussed in Sec.\,\ref{Sec: Pump sources}. 

To conclude, it should be noted that we have greatly simplified the description of the photoemission process. In describing the final state as a free electron, we have neglected the presence of scattering states, which can affect the photoemission matrix element \cite{Gierz2011}. We have also invoked the sudden approximation and the three-step model of photoemission, which may break down in the low kinetic energy limit \cite{Gadzuk1975}. The photoemission process itself is being studied using attosecond metrology techniques such as streaking and RABBITT interferometry \cite{Ossiander2018}. Though beyond the scope of this review, the timing of the initial excitation of the electron wavepacket, transport, and emission will inevitably lead to more advanced modelling of the photoemission process.

\begin{figure}[t!]
    \centering
    \includegraphics{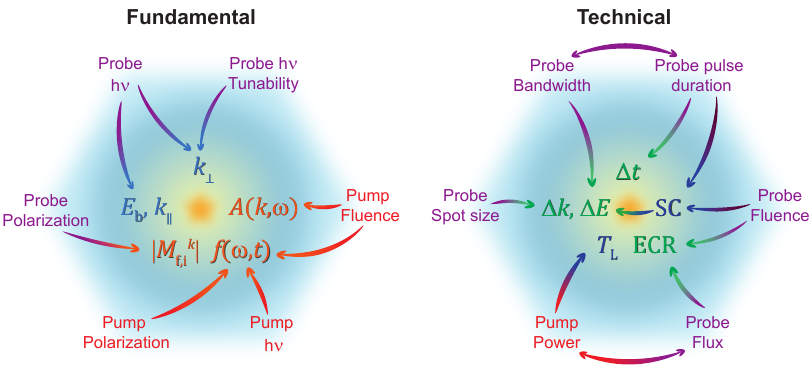}
    \caption{\textbf{Intertwined properties of the source (pump and probe) parameters and fundamental and technical aspects of photoemission.} In both panels probe (pump) parameters are shown outside in purple (red). Left: The fundamental aspects of photoemission related to these parameters. The electron binding energy, crystal momentum (blue), and ARPES intensity (orange) are shown inside. Arrows illustrate how laser parameters affect these terms in photoemission. Right: Technical aspects of photoemission are shown on the inside. Energy, angle, and time resolution are given by $\Delta E$, $\Delta k$ and $\Delta t$, respectively. The lattice temperature on the sample is given by $T_\text{L}$. SC: space charge, ECR: Electron-count rate. Quantities to optimize are shown in green, while limitations are shown in dark blue. Trade-offs between laser parameters are discussed in the text.}
    \label{Fig: Laser_params}
\end{figure}

\section{Technical aspects of TR-ARPES}
\label{Sec: Technical}
In the previous section, we discussed how the pump and probe properties couple to fundamental aspects of the photoemission process. These properties are summarized in the left side of Fig.\,\ref{Fig: Laser_params} with probe and pump parameters in purple and red, respectively. The energy $E_\text{b}$ and momentum ($\boldsymbol{k}_\parallel, k_\perp$) are given in blue. The matrix element $|M_{f,i}^{\boldsymbol{k}}|$, occupation function $f(\omega, t)$, and spectral function $A(\boldsymbol{k},\omega)$, which contribute to the ARPES intensity are given in brown.

In any TR-ARPES measurement, the recorded spectrum can differ from fundamental expectations for a variety of technical reasons. A summary of these technical considerations and relevant source parameters is shown on the right side of Fig. \ref{Fig: Laser_params}. In this section, we focus on three key technical considerations that are directly affected by source parameters: (i) time and energy resolution, (ii) space charge (SC) distortion, and (iii) thermal effects. Parameters we wish to optimize are shown in green, namely: system resolutions ($\Delta k$, $\Delta E$, $\Delta t$) and electron count rate (ECR), while parameters that force us to compromise are shown in dark blue, namely: thermal effects from the pump and SC distortion. Ultimately, no single set of source parameters can provide the optimal conditions for all TR-ARPES experiments, so the tunability of source parameters is important. In this respect, some parameters such as fluence or polarization may be easier to adjust than others, such as repetition rate and pulse duration. Lastly, as foreshadowed in Sec.\,\ref{Sec: History}, the interplay between electron spectrometers and the laser source must also be considered. Although a complete discussion of the hemispherical, time-of-flight and momentum-microscope analyzers is beyond the scope of this paper, we highlight key advantages and disadvantages of each as it relates to our technical considerations of the laser source.

\subsection{Trade-offs: System resolutions}
The energy and momentum resolution of the ARPES system ($\Delta E, \Delta k$) broadens the measured spectra, as seen in Fig.\,\ref{Fig: specfun}. The system resolution relies on the electron spectrometer $\Delta E_\text{s}$ and the bandwidth of the probe light source $\Delta E_\text{p}$. For equilibrium ARPES, light sources such as gas-discharge lamps and continuous wave (CW) or quasi-CW lasers can produce light with narrow spectral bandwidths, such that $\Delta E_\text{p}\leq \Delta E_\text{s}$. Other sources, such as synchrotron radiation, rely on monochromators to control the spectral bandwidth, which may be larger than the detector resolution at high photon energies.

Of the modern analyzers, the hemispherical analyzer (HA) is the most common and is capable of achieving energy and momentum resolution of $\Delta E_\text{s}^\text{HA}\approx 1$~meV and $\Delta\theta^\text{HA}<0.1~^{\circ}$ \cite{Sobota2021}. However, lens modes and entrance slit sizes impact the resolution achieved. Operating at the minimum resolution of a HA requires a small entrance slit, leading to a significant reduction of the ECR. Such a severe trade-off is often undesirable, and the analyzer is operated with a resolution near 10~meV. Angle-resolved time-of-flight spectrometers discriminate the electron kinetic energy with a delay-line detector (DLD) and have demonstrated an energy and angular resolution similar to that typically achieved by HAs at low photon energies ($\approx 10$~eV) \cite{Berntsen2011}. However, $\Delta E_\text{s}^\text{ARTOF}$ is dependent on the time resolution of the DLD and, by extension, on the kinetic energy of the photoelectron \citep{Kuhn2018}. Lastly, the momentum-microscope (MM)-ARTOFs have demonstrated similar resolution ($\Delta E_\text{s}^\text{MM}\approx 10$~meV), but achieving this resolution depends on minimizing space charge effects, as we discuss below.

The contribution to the energy resolution from the probe bandwidth varies from $10-500$~meV and is dependent on the generation process. In most cases, this is larger than the analyzer contribution. The large probe bandwidth directly results from using femtosecond pulses, in contrast to the CW or quasi-CW ($\gtrapprox$ 10~ps pulses) employed in equilibrium ARPES. The Fourier transform-limited duration of the probe pulse, which contributes to the time resolution $\Delta t$ of the system, is inversely related to its bandwidth. Hence, $\Delta E_p$ and $\Delta t$ are fundamentally limited by each other and must be considered together. From a scientific perspective, the balance of bandwidth and pulse duration is driven by the physics being studied. However, the connection between pulse parameters and the ``accessible physics" is not straightforward. The same physics can manifest in markedly different ways for different material systems and pump excitation regimes. For instance, the electron-phonon interaction may manifest as a kink in the band dispersion \cite{Zhang2014}, oscillations of the electronic band structure \cite{Gerber2017}, non-thermal occupation of electrons \cite{Na2019}, and the timescales of the electronic temperature decay \cite{Perfetti2007, Yang2015}. Therefore, one must consider the spectral features of interest for each specific case. Moreover, this trade-off must be considered from the technical perspective of light generation. For instance, the generation of UV photons from frequency up-conversion in solids offers greater flexibility than the VUV and XUV photons from high harmonic generation in gases, which require much higher peak electric fields (and hence shorter pulses) for efficient generation \cite{Gauthier2020}. Ultimately, the Fourier transform-limited time-bandwidth product is the best one can do. A detailed discussion of different sources and their approach to the $\Delta E$/$\Delta t$ trade-off is given in Sec.\,\ref{Sec: Laser survey}.

\begin{figure}[t!]
    \centering
    \includegraphics{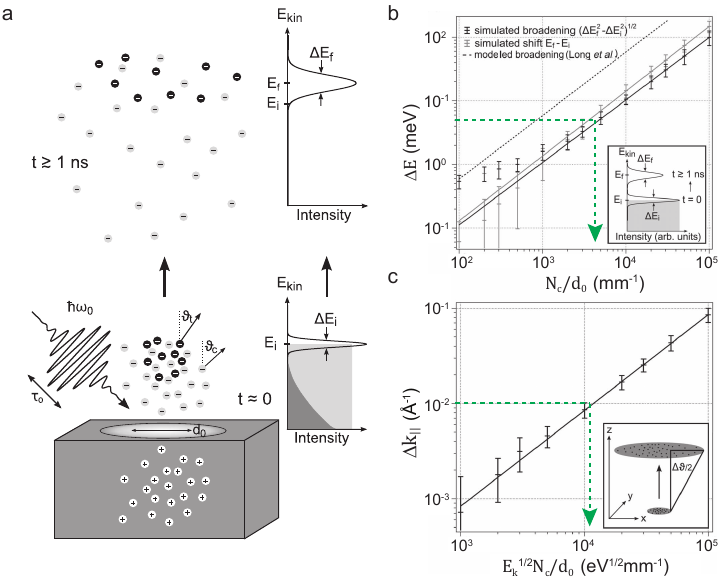}
    \caption{\textbf{Simulation of space charge (SC) effects in valence band spectroscopy.} Figure adapted from Hellmann \textit{et al.} \cite{Hellmann2009}. \textbf{(a)} Sketch of the SC effect. A probe pulse of photon energy $\hbar\omega_0$ and duration $\tau_0$ is focused onto the sample with spot size $d_0$, generating photoelectrons. The photoelectron kinetic energy and emission angle change as they propagate due to SC. A Gaussian spectrum of ``test" electrons (black) defined by position $E_i$ and width $\Delta E_i$ is selected. Cloud electrons are emitted isotropically. Changes in the test spectrum due to the cloud electrons (light grey), mirror charges (white), and the solid (dark grey) are simulated. \textbf{(b)} Simulations of the energy shift and broadening of valence band electrons as a function of photoelectron cloud density $N_c/d_0$. The error bars show the variation of parameters $d_0~(0.04, 0.1, 0.4)$~mm and $E_i~(1, 10, 100)$~eV, and $\tau_0$ (10~fs, 10~ps). Inset: Initial and final configurations. Test electrons have $\Delta E_i = 5$~meV and $\theta_t=0$. \textbf{(c)} Simulation of the momentum broadening using the same parameters in (b), except that $E_i (1, 25, 100)$~eV are used. Inset: Real-space divergence of photoelectrons. Green arrows indicate ideal limits in SC-induced broadening.}
    \label{Fig: SC}
\end{figure}

\subsection{Trade-offs: Space charge}
\label{Sec: space charge}
In TR-ARPES, space charge (SC) can negatively impact the energy and momentum resolution. The SC distortion originates from the Coulomb repulsion between electrons photo-emitted near each other in space and time, which leads to a broadening and shift of the photoelectron kinetic energy and angular distribution \cite{Zhou2005}. In experiments performed with quasi-CW pulsed sources, the photoemission events are spaced far enough apart in time that the Coulomb repulsion between fast-moving photoelectrons is weak ($E_\text{k}=2.84$~eV corresponds to $v_e\sim 10^6$~m/s). For experiments performed with ultrafast pulses, multiple photoemission events occur within a window given by the probe pulse duration ($\tau_0\approx 100$~fs), and SC is a significant issue \cite{Hellmann2009,Zhou2005, Passlack2006, Mathias2010}. As TR-ARPES sources improve in energy resolution, analysis of the photoexcited electronic dispersion, photo-induced phase transitions and associated gap openings, and evolution of the electronic self-energy are within reach. Space charge can significantly reduce the fidelity of these detailed analyses, and it is imperative that its effects are minimized. 

A simulation of the SC effect adapted from Hellmann \textit{et al.} is given in Fig.\,\ref{Fig: SC}a \cite{Hellmann2009}. In valence-band photoemission spectroscopy, SC typically manifests as an upward shift of the Fermi energy and a broadening in photoelectron kinetic energy and angular distribution. The broadening and shift are proportional to the number of photoelectrons ($N_c$) generated per pulse, and inversely proportional to the pulse duration ($\tau_0$) and the lateral spot dimension ($d_0$). These dependencies are explicit in Fig.\,\ref{Fig: SC}b and for a range of pulse parameters typically encountered in TR-ARPES: $d_0~(0.04, 0.1, 0.4)$~mm, $E_i~(1, 10, 100)$~eV, and $\tau_0$~(10~fs, 10~ps). The green dashed line in Fig.\,8b indicates that $N_c/d_0$ must be less than $3000$~mm$^{-1}$ per pulse to limit the broadening to less than 5~meV, which is a typical acceptable value for TR-ARPES setups with an overall energy resolution on the order of 10-100~meV. The angular/momentum broadening is given in Fig.\,8c. Relative to the momentum resolution of modern analyzers (green dashed line, $<$0.01\AA$^{-1}$), the constraint on the photoelectron density from the angular broadening is $\sqrt{E_k}N_c/d_0 < 10000$~eV$^{1/2}$mm$^{-1}$ per pulse, slightly less stringent than that of the energy broadening \cite{Hellmann2009}. This SC simulation provides an important order-of-magnitude estimation in TR-ARPES studies of valence bands. However, the relation between probe fluence and photoelectron density will vary for different materials. In addition to probe-induced SC, one must also ensure that the pump fluence remains low enough not to generate low kinetic energy photoelectrons through multi-photon photoemission and plasmonic emission from metallic surface defects. The pump-induced photoelectron cloud has much lower kinetic energy. Near the sample, it contributes a delay-dependent component to the SC-induced energy shift that is difficult to model \cite{Oloff2016, Kutnyakhov2020}. 

\begin{figure}[t!]
  \begin{center}
    \includegraphics{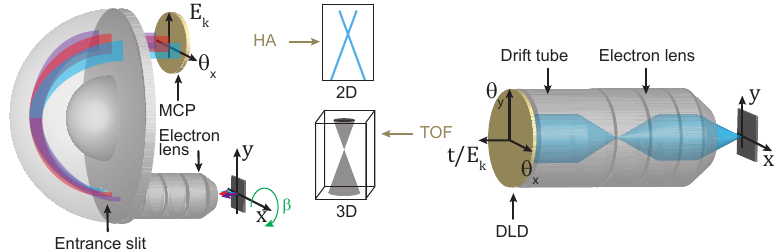}
    \caption{\textbf{Comparison of electron spectrometers.} Left: The hemispherical analyzer (HA): photoelectrons are collected by the electrostatic lens. Only electrons parallel to the entrance slit are transmitted to the HA. A voltage between the two concentric hemispheres discriminates $E_\text{k}$ on the axis perpendicular to the entrance slit. The two-dimensional micro-channel plate (MCP) encodes $E_\text{k}$ and $\theta_x$, generating a 2D dispersion. $\theta_y$ is obtained by rotating the sample through $\beta$ (green). Right: The angle-resolved time-of-flight spectrometer (ARTOF): As in the HA, a portion of the total photoelectron distribution (solid blue) is collected by the electrostatic lens. Without an entrance slit, $\theta_x$ and $\theta_y$ are collected. The photoelectrons then travel through the field-free drift tube. The two-dimensional MCP maps $\theta_x$ and $\theta_y$ simultaneously, while $E_\text{k}$ is discriminated by the delay-line detector (DLD) based on the arrival time $t$.}
  \label{Fig: detector}  
  \end{center}
\end{figure}

Reducing the probe fluence by increasing its spot size is one way to minimize space charge. However, small spot sizes ($\lesssim 100~\upmu$m) are needed to probe uniform (single-crystal) domains. In practice, the probe flux is reduced to compensate for SC distortions. The consequence is a reduction of the ECR and an increase in acquisition time, which can quickly scale beyond reasonable limits. The ECR also depends on the throughput of the electron spectrometer. Here, the ARTOF may be advantageous over the HA as it can capture the entire photoelectron distribution $(E_\text{k}, \theta_x, \theta_y)$ generated per pulse. In contrast, the HA captures only one cut of the distribution $(E_\text{k}, \theta_x)$ (see Fig.\,\ref{Fig: detector})\cite{Kirchmann2008}. However, this parallel detection is only time-saving if the ARTOF can operate with a photocurrent comparable to that of the HA. 
\begin{figure}[t!]
  \begin{center}
    \includegraphics{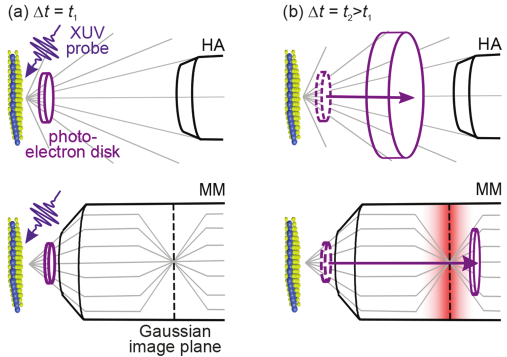}
    \caption[Space charge in the momentum microscope.]{\textbf{Space charge in the momentum microscope (MM).} Figure  adapted from Maklar \textit{et al.} \cite{Maklar2020}. \textbf{(a)} Schematic of the photoelectron disk at $t_1$ just after generation by an XUV pulse. \textbf{(b)} The photoelectron disk at $t_2$ sometime after generation. In the HA (top), the electron disk spreads over the complete $2\pi$ solid angle and broadens along the direction of propagation. In the MM-ARTOF (bottom), the large acceleration voltage of the MM guides all photoelectrons into the lens column. After acceleration, the relative kinetic energy difference between electrons is small, leading to a thin dense disk. Space charge is exacerbated at the focal planes of the MM (red).}
  \label{Fig: MM}  
  \end{center}
\end{figure}

A recent development in TR-ARPES is the integration of the momentum microscope lens column with the ARTOF (MM-ARTOF)  \cite{Madeo2020, Schonhense2021, Schmitt2022, Maklar2020}. In the MM, a large positive voltage is applied to the front of the lens column, which allows the entire photoelectron distribution to be collected. Apertures in the image plane can reduce the region of the sample from which photoelectrons are collected, irrespective of the spot size of the probe. The electron lens then transforms the distribution back into the reciprocal image plane. The MM seems ideal for avoiding space charge; the extraction voltage collects all photoemitted electrons, and the image-plane filtering allows one to increase the probe spot size without consequence. A quantitative comparison between a HA and an MM-ARTOF has been performed by Maklar \textit{et al.} \cite{Maklar2020}. This comparison shows that while the MM is powerful, it is particularly susceptible to space charge inside the lens column when operating in its spatially-filtering lens mode (see Fig.\,\ref{Fig: MM}). In the HA, the photoelectron distribution expands as it propagates due to the difference in the kinetic energy. The high acceleration voltage in the MM reduces the relative kinetic energy difference, creating a high spatial density of photoelectrons in various focal planes. The high electron density leads to SC-induced broadening. Future technical development will likely improve its performance and capabilities for detailed spectroscopic studies \cite{Schonhense2021b}.

The last way to increase the ECR while avoiding space charge is by increasing the repetition rate, which increases the photocurrent while keeping $N_c$ below SC limits \cite{Sobota2012, Peli2020, Mills2019, Corder2018, Ozawa2015, Bao2021}. However, the repetition rate also affects other experimental components. For instance, the DLD detection scheme employed in ARTOFs requires a repetition rate typically $\leq 1$~MHz to prevent the mixing of fast electrons from one wavepacket with slow electrons in the previous wavepacket. In this respect, the recent integration of high-pass energy filtering into TR-ARPES has enabled ARTOFs to operate at higher repetition rates \cite{Kunin2023, Wallauer2021}. Several aspects of TR-ARPES source design also become more challenging at high repetition rates due to the lower pulse energy available for the nonlinear frequency conversion process (these are discussed in Sec.~\ref{Sec: Probe sources}). Finally, there is the unavoidable issue of the increase in pump photon flux and the associated average-power heating of the sample at high repetition rates.

\subsection{Trade-offs: Thermal effects}
Thermal dissipation in ultrafast experiments depends strongly on the material as well as the characteristics of the laser pulse, such as average power and fluence. Hence, the thermal damage limit has properties that are both fundamental and technical in origin. At low repetition rates, the pump energy is typically dissipated before the next pulse, though some exceptional cases exist \cite{Crepaldi2022}. If the system fully relaxes between pulses, then only the fundamental processes involved in pump excitation are important. At higher repetition rates, the thermal dissipation may not be sufficient to return the system to equilibrium between pulses. In this case, sample damage or the maximum desired sample temperature establishes an upper limit on the average pump power used at a given repetition rate. This limitation is technical in that the steady state temperature is dependent on the thickness of a sample, which can be reduced substantially, even as far as a single-atomic layer. Even so, experiments performed at high repetition rates will not reach the highest pump fluence regimes due to limitations on the lattice temperature $T_\text{L}$. In Fig.\,\ref{Fig: Laser_params}, the repetition rate connects the pump power and probe flux parameters, and the trade-off of these parameters leads to a trade-off between $T_\text{L}$ and ECR.

One can increase the pump fluence without increasing the power by decreasing the pump spot size. In TR-ARPES, the pump spot size is typically kept larger than the probe, such that a uniformly photoexcited region is measured. If the probe is kept large to mitigate SC and the pump is kept small to increase fluence, then another way to restrict the measured region is needed. Here, the spatial filtering capability provided by the momentum microscope is highly advantageous. It has been used in the study of exfoliated materials \cite{Madeo2020} and heterostructures \cite{Schmitt2022} using HHG sources.

In summary, we discussed how light source parameters are deeply intertwined through different technical aspects of TR-ARPES experiments and how developments in electron spectrometers are tackling the resulting trade-offs on several fronts. The two key takeaways from this section are: (i) the inverse relationship between pulse duration and bandwidth leads to a trade-off between the energy and time resolution, which affect the kind of timescales and spectral features we can study; (ii) space charge is a pervasive hindrance in time-resolved studies. Efforts to avoid it while keeping ECR high motivate high repetition rates sources, but the corresponding high average pump power causes thermal issues on the sample. Accounting for thermal effects may place a limit on the maximum pump fluence.

\section{TR-ARPES laser sources: Recent trends}
\label{Sec: Laser survey}
\begin{figure}[t!]
  \begin{center}
    \includegraphics{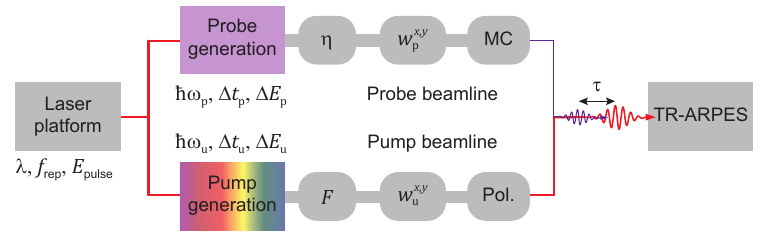}
    \caption{\textbf{Block diagram of TR-ARPES components considered in Sec.\,\ref{Sec: Laser survey}.} A laser platform (Ti:sapphire or Yb-doped) is chosen, with characteristic fundamental frequency/wavelength $\lambda$, repetition rate $f_\text{rep}$, and pulse energy $E_\text{pulse}$. The pulses seed both pump and probe generation. Here, the range of photon energies for pump ($\hbar\omega_\text{u}$) and probe ($\hbar\omega_\text{p}$), the bandwidth ($\Delta E_\text{p}$, $\Delta E_\text{u}$) and the pulse duration ($\Delta t_\text{p}$, $\Delta t_\text{u}$) are important considerations. Lastly, the ``beamline" components manipulate the properties of the pump and probe and include elements for spectral separation (MC: monochromator), polarization (Pol.) and fluence ($F$) control, and spot size ($w^{x,y}$), among others. Optics used in the beamline must maintain a high efficiency $\eta$ and low dispersion.}
  \label{Fig: Block}  
  \end{center}
\end{figure}

\begin{figure}[t!]
    \centering
    \includegraphics{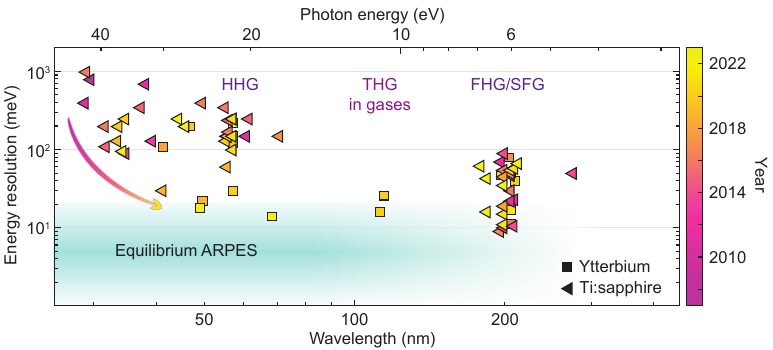}
    \caption{\textbf{Survey of probe sources used in TR-ARPES experiments.}  The energy resolution $\Delta E$ determined from a Fermi-edge fit on a polycrystalline metallic sample (usually Au) is shown versus photon energy. Frequency up-conversion in solids generates photons in the 4.5-7~eV range and are continually developed over the years, highlighting the reliability and relevance of the system. UV-driven third-harmonic generation (THG) in gases pushes the photon energy up to 11~eV and maintains a good energy resolution but is sparsely developed. High-harmonic sources (HHG), particularly those based on Yb-doped lasers, have demonstrated a huge improvement in energy resolution in recent years (gradient arrow). The references in this figure can be found in Tab.\,\ref{Tab: probe_survey}.}
    \label{Fig: Probe_survey}
\end{figure}

In this section, we explore the trade-offs of Sec.\,\ref{Sec: Technical} in the context of the technical construction of the experimental setup. The components we consider are sketched in Fig.\,\ref{Fig: Block}. We begin with a comparison of the Ti:sapphire and Yb-doped laser platforms, followed by pump and probe generation and beamline components. As a part of this discussion, we survey existing probe sources from ``technical" publications devoted to instrumentation development, as they report technical advances that boost the capabilities of TR-ARPES. For clarity, we include a comprehensive list of parameters and citations in the \ref{Sec: Tables}, which are plotted in Figs. \ref{Fig: Probe_survey} and \ref{Fig: Probe_param}. Using this survey as a framework, we discuss key aspects of the technical development that a system designer should consider when choosing a laser platform and constructing an apparatus for TR-ARPES. Much of this discussion centers around recent developments in pump and probe technology, including (i) improving probe energy resolution; (ii) expanding pump wavelength tuning range; (iii) optimizing the time-bandwidth product; (iv) achieving small spot size on the sample; and (v) adding polarization control of the probe. For the most part, we limit our discussion to laser systems applied directly in TR-ARPES experiments, and we include a very limited discussion of equilibrium ARPES and technical non-ARPES laser systems as appropriate. We encourage the reader to explore the research cited in these references as needed.

\subsection{Laser platforms in TR-ARPES}
Over the past three decades, researchers have continually leveraged state-of-the-art laser technology when developing laser sources used in ARPES and TR-ARPES. As illustrated in Fig. \ref{Fig: history}, this progression has moved from ultrafast dye lasers to more user-friendly and technically capable solid-state laser systems. Today, lab-based TR-ARPES systems are generally based on three specific gain media:  Ti:sapphire, Yb-doped fiber, and Yb-doped crystals. 

The front end of laser systems are usually mode-locked oscillators, which generate femtosecond-duration pulses at repetition rates of 40-100~MHz. Variable repetition rate sources can be produced with the use of a pulse-picker at the output of the oscillator or inside the cavity in a `cavity dumper' configuration. Generally, a low-noise oscillator is desirable for the long-term stability of pulse shape, pulse energy/average power, and beam pointing. In some applications, such as few-cycle pulse systems or cavity-enhanced HHG systems, the carrier-envelope offset phase stability of the oscillator is important, which requires care in the design. The timing jitter is another factor that might influence the choice of the mode-locked oscillator or mode-locking regime, particularly if the pump and probe are derived from separate oscillators. Among the high-power commercial sources, simultaneous control of carrier-envelope offset phase and repetition rate stability is usually not standard but is available as an option in some cases.

The laser oscillator output is typically amplified externally to produce a higher pulse energy. Amplifier systems can be based on multiple passes through the gain medium or in a cavity-based regenerative amplifier, where a laser pulse is coupled into a resonator and undergoes several round trips through the gain medium before being coupled out. One major advantage in reducing the repetition rate in laser systems is that higher energy pulses can be generated because the pump energy is converted to fewer pulses in the laser or amplifier gain medium. In high-power systems, tuning the repetition rate after manufacturing is not always possible due in part to the nonlinearities in pulse propagation that can arise with the different pulse energies generated at vastly different repetition rates.  

The early proliferation of TR-ARPES systems followed the developments of Ti:sapphire lasers as they progressed towards higher repetition rate, higher average power, and shorter pulse duration. Typically operated in the 780~nm to 840~nm wavelength range (1.6~eV to 1.47~eV), Ti:sapphire laser outputs can be up-converted with nonlinear crystals, such as beta-barium borate (BBO). Photon energies (4.5-5.2~eV) produced by third-harmonic generation (THG) or sum-frequency generation (SFG) can be used to study the conduction band dynamics of semiconductors, while the 4th harmonic (5.9~eV - 6.35~eV) can be used to study dynamics of the valence and conduction band. Ti:sapphire lasers are ideal for 6-eV sources, experiments with high time resolution, and HHG at lower repetition rates, yet it is very challenging to increase the average power sufficiently to perform HHG at repetition rates of 100 kHz or higher.

Conversely, higher average power can be achieved with Yb-doped lasers and amplifiers --operating near 1030~nm or 1.19~eV-- but the gain bandwidth and wavelength tuning range is not as large as in the Ti:sapphire. As a result, when operating in the linear amplification regime, pulses are limited to approximately (transform-limited) 200-300~fs. Shorter pulses can be produced using nonlinear amplifiers or through nonlinear spectral broadening in optical fibers or gas-filled Herriott cells. However, careful management of both the nonlinearity and dispersion is required; very often, these pulses are not transform-limited. At its fundamental energy of 1.19~eV, overcoming the work function requires the photon energy of the fifth harmonic to be generated via sequential stages of nonlinear conversion and sum frequency mixing \cite{Boschini2014, Ishida2016}. Alternatively, with the average power afforded by new Yb-doped laser sources, it is also possible to use the second harmonic of the 1.19 eV fundamental to pump an OPA to provide tunable radiation near 780 nm, which can be converted to the UV via fourth harmonic generation \cite{Kremer2021}. While frequency up-conversion in solids based on Yb-doped crystals and fibers is much less common than that in Ti:sapphire sources, the power scaling of Yb-doped laser sources can outperform Ti:sapphire lasers by one to two orders of magnitude, making it attractive for seeding extremely nonlinear frequency up-conversions such as UV-driven THG in gases and HHG. 

\subsection{Probe sources}
\label{Sec: Probe sources}
The generation of TR-ARPES probe beams is performed via nonlinear frequency up-conversion of the laser source. The first metrics considered for the probe are typically the photon energy range and energy resolution. The energy resolution of these sources --typically extracted from a measure of the Fermi-edge at cold temperatures on a polycrystalline metal-- is shown in Fig.\,\ref{Fig: Probe_survey} versus photon energy. The colour of the marker indicates the year in which the paper was published, and the shape of the marker denotes the gain medium of the source. For clarity, we will use $n$H to refer to the $n^{\text{th}}$ harmonic of the laser fundamental frequency (FF). 
Up to 6-7~eV, the probe beam is usually generated from the FF via frequency mixing in nonlinear crystals using third harmonic generation (THG), fourth harmonic generation (FHG) or sum frequency generation (SFG) between the various components. THG in xenon gas (driven by the 3H) can push the probe up to as high as 11~eV, and HHG is used to produce higher photon energies. These three classes of probe generation have significantly different technological requirements, which we discuss in the following subsections.

The energy resolution of typical equilibrium ARPES experiments ($\leq10$~meV) is shown in light blue at the bottom of Fig.\,\ref{Fig: Probe_survey}. Of the sources surveyed, 6-eV sources have the highest energy resolution. Using a portion of the fundamental as a pump, these systems show impressive flexibility despite the limited momentum space coverage and have produced many scientific works studying topics such as electron-boson coupling in cuprates \cite{Smallwood2012, Perfetti2007, Graf2011}, charge-density waves in transition-metal dichalcogenides (TMDs) \cite{Hellmann2012, Schmitt2008, Petersen2011}, topological insulators \cite{Sobota2012, Wang2012, Crepaldi2012}, and more \cite{Perfetti2008}. Due to their accessibility and reliability, 6-eV sources are a staple in TR-ARPES labs and continue to be built every year. 

Sources based on THG and HHG did not proliferate at the same rate as 6-eV sources with the advent of commercial Ti:sapphire lasers due to their technical complexity and high pulse energy requirement. The desire to study spectral features at the BZ edge of materials such as graphene and TMDs spurred the development of HHG-based sources \cite{Mathias2007}. Due to limited average power, Ti:sapphire-based systems operated at kilohertz repetition rates and utilized short pulses to generate the high peak electric fields needed for efficient HHG. The resulting large bandwidth and poor energy resolution are on the order of several hundred meV \cite{Rohwer2011, Frassetto2011, Heyl2012}. Nevertheless, the high time resolution enabled studies of electron-electron interactions and carrier dynamics in graphene \cite{Gierz2013, Ulstrup2015}, graphite \cite{Stange2015} and TMDs \cite{Cabo2015, Wallauer2016}. In recent years, the energy resolution of HHG sources has dramatically improved and is competitive with modern 6-eV systems. We observe from Fig.\,\ref{Fig: Probe_survey} these are largely Yb-doped sources. The available power scaling of Yb-doped fiber amplifiers produces higher pulse energies with longer pulses and narrower bandwidth. Other approaches to increasing the high-harmonic yield at longer pulses and narrower bandwidths are discussed in detail in Sec.\,\ref{Sec: HHG}.

\begin{figure}[t!]
    \centering
    \includegraphics{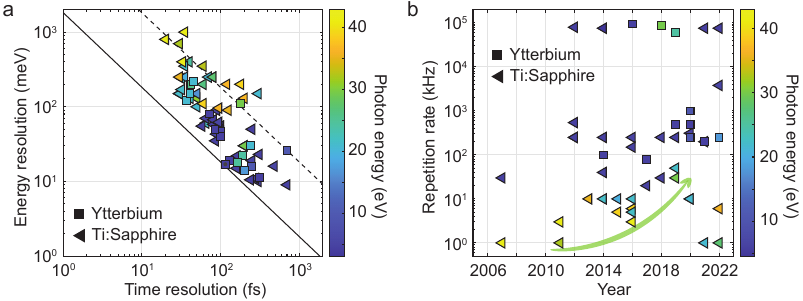}
    \caption{\textbf{Survey of probe sources: time-bandwidth product and repetition rate.} \textbf{(a)} Energy and time resolution of TR-ARPES probe sources. The energy resolution $\Delta E$ is typically measured from the Fermi-edge of a polycrystalline sample. The time resolution $\Delta t$ is defined as the pump-probe cross-correlation. The solid line represents the Gaussian transform limited time-bandwidth product at 1825~meV$\cdot$fs. The dashed line represents 10$\times$1825~meV$\cdot$fs. Recent efforts have dramatically improved the time-bandwidth product of XUV probes to be on par with that of 6-eV probes. \textbf{(b)} Repetition rate of TR-ARPES sources as a function of year. Ti:sapphire sources from 10~kHz to 10~MHz regime have generally been available for generating 6-eV. In contrast, Ti:sapphire-based HHG sources show an increasing trend from 1~KHz to 100~kHz. Ytterbium-doped sources pushed the repetition rate of HHG past the 100~kHz limit.}
    \label{Fig: Probe_param}
\end{figure}

The trade-off of time and energy resolution of the probe sources is shown in Fig.\,\ref{Fig: Probe_param}a, with the solid line defining the transform-limited time-bandwidth product. The energy resolution $\Delta E$ is typically obtained from a fit of the Fermi-edge of a polycrystalline sample (Au) measured at low temperature. The time resolution $\Delta t$ is defined by the cross-correlation between pump and probe pulses. Reaching the transform-limited time-bandwidth product is not trivial and usually requires optimization at every stage of the probe generation and beamline. We see from Fig.\,\ref{Fig: Probe_param}a that Ti:sapphire and Yb-doped 6-eV systems come closest to this transform-limited performance. The 6-eV sources have not only the highest energy resolution but also the highest tunability in the time-bandwidth product, allowing the user to tailor $\Delta E$ and $\Delta t$ based on the physics to be studied \cite{Gauthier2020}. The performance of HHG sources around 25~eV is also fairly reasonable, with a few setups demonstrating the same performance as that of 6-eV sources. Photon energies above 30~eV are less optimized and have $\Delta E >100$~meV.

Lastly, the repetition rate of probe sources is shown as a function of the publication year in Fig.\,\ref{Fig: Probe_param}b. In the modern history of TR-ARPES, Ti:sapphire oscillator systems with repetition rates from 10~kHz to $>10$~MHz have always been available with high enough pulse energies for efficient frequency up-conversion into the UV (6~eV) range. In contrast, the development of HHG sources shows a clear trend of steadily increasing repetition rates --to increase the ECR-- while maintaining high enough pulse energies for an efficient harmonic generation. In this progression, Ti:sapphire-based systems were the only sources up until 2018, with the highest repetition rates in the 100~kHz regime. Recently developed Yb-doped sources using commercial lasers have emerged in the 100~kHz to 1~MHz regime, and cavity-enhanced HHG sources operating at nearly 100~MHz using Yb-doped fiber lasers have been successfully integrated into TR-ARPES measurements.

As a result of these trade-offs, all of these laser platforms find their place in applications. The past several years have seen commercial lasers mature dramatically, evidenced by the increase in the availability of sources offering turn-key operation and little to no user intervention. Today's commercial lasers achieving lab-tested long-term reliability are available from multiple vendors at a variety of repetition rates, and they readily produce 10-100~W average power, and even more in some cases.

\subsubsection{Nonlinear crystal based sources: \texorpdfstring{$\chi^{(2)}$}{X(2) } processes in solids}
\label{Sec: cyrstal NLO}
Nonlinear frequency conversion via the second-order polarization, $\chi^{(2)}$, in solids is the most straightforward means of generating a probe with sufficient photon energy for photoemission. A commercial Ti:sapphire laser and two or three stages of SHG/SFG are all that is needed to get started. From this starting point, numerous system refinements can be made, such as: (i) pushing the limits of probe photon energy in nonlinear crystals; (ii) optimizing the balance between time and energy resolution of both the pump and probe pulses; and (iii) minimizing the spot size on the sample for $\upmu$ARPES studies (details in Sec.\,\ref{Sec: smallspot}).

\begin{figure}[t!]
    \centering
    \includegraphics{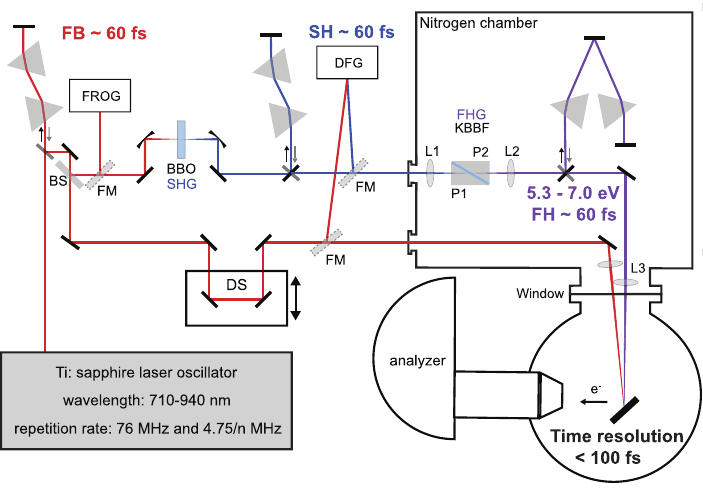}
    \caption{\textbf{Experimental setup of a tunable VUV source using KBBF crystals.} Figure adapted from Zhong \textit{et al.} \cite{Zhong2022}. BS: beam splitter, DFG: difference frequency generation, DS: delay stage, FB: Fundamental beam, FH(G): fourth harmonic (generation), FM: Flip mirror, FROG: Frequency-resolved optical gating, SH(G): second harmonic (generation). The KBBF device has a crystal thickness of 0.06~mm and small coupling prisms of length $L=7.5$~mm. A previous device from Bao \textit{et al.} has a 1~mm thick crystal and prisms with $L=16.5$~mm \cite{Bao2022}.}
    \label{Fig: KBBF}
\end{figure}
The transparency window of the nonlinear medium is the ultimate limit for crystal-based sources. However, it is usually not possible to produce light via cascaded SHG to this limit because the birefringence limits the phase-matching range. SFG between the fundamental and third-harmonic (i.e. $\omega + 3\omega$) can usually be phase-matched at higher photon energies than SHG and has been exploited in numerous 6-eV sources \cite{Faure2012}. For 6-eV sources, $\beta$-barium borate (BBO) is the most commonly used. It has a transparency cut-off wavelength of 189~nm (6.56~eV), and while SHG can only be performed to about 205~nm (6.05~eV), a $3\omega+\omega$ wave-mixing scheme can be phase-matched near 189~nm. Lithium triborate (LBO) has an attractive transparency cut-off near 158~nm (7.85~eV) but a birefringence that allows phase-matching down to only 243~nm (5.1~eV). A review of this property for several nonlinear crystal materials is summarized in \cite{Zhou2018}.

Potassium beryllium fluoroborate (KBBF) is a crystal that has received recent attention for its ability to push the photon energy above 7~eV. Its implementation is more technically challenging, partly due to absorption in the air at higher photon energy. The typical geometry used for KBBF crystals uses a thin crystal mounted between two prisms to permit phase-matching and transmission of the generated wave out of the crystal \cite{Zhang2009}. This geometry has enabled KBBF to be used in high-resolution laser ARPES \cite{Kiss2008}, but the prism dispersion must be accounted for with ultrashort pulses. The experimental setup from Zhong \textit{et al.} is shown in Fig.\,\ref{Fig: KBBF} \cite{Zhong2022}. With chirp pre-compensation of the second harmonic of a Ti:sapphire laser and compression of the generated fourth harmonic, this KBBF setup achieves a time resolution of 280~fs \cite{Bao2022}. Very recently, a thin KBBF crystal and reduced-thickness prism device helped achieve a time resolution of 81-95~fs in TR-ARPES measurements with a probe tuning range of 5.3-7.0~eV \cite{Zhong2022}.

In Fig.\,\ref{Fig: Probe_param}a, we see nonlinear crystal-based sources come closest to the Gaussian transform limit. The ability to access different time and energy resolutions is desirable for tailoring the experiment to the relevant spectral features and dynamics of each material. Hence, the tunability of the pulse duration and bandwidth at this limit is a subject of interest. In this respect, the wavelength-dependent refractive index (i.e. dispersion) of nonlinear crystals affects the bandwidth of the generated wave by limiting the bandwidth that can be phase-matched. Generally, one must match the acceptance bandwidth of the crystal to the input pulse bandwidth, and the compression of the input/output pulses must be optimized to produce the shortest 6-eV pulses. The length of the crystal can be increased to narrow the bandwidth of frequency-converted pulses. This provides a useful tuning mechanism, although it does not necessarily maintain the minimum time-bandwidth product. A detailed analysis of this problem in 6-eV generation via $2\omega + 2\omega$ mixing in BBO with Gaussian pulses is presented in \cite{Gauthier2020}. For pulses that are not Gaussian, the situation is likely less straightforward. Nevertheless, tuning the pulse bandwidth by changing the length of nonlinear crystals is the easiest means of controlling the time-bandwidth properties of 6-eV sources.

The beamline requirements for a 6-eV system are the simplest of the TR-ARPES probes as the beam can travel through the air and enter the vacuum chamber through a CaF$_2$ or MgF$_2$ viewport. The 7-eV sources, however, require vacuum or nitrogen environments, which add complexity. Optics for polarization control in the range of 6~eV are also available commercially. In principle, the short wavelength helps to ease the requirements on focusing in typical UHV vacuum chambers with relatively long working distance from a focusing lens to the sample. A large aperture design of the entrance window for the 6~eV light into the vacuum system helps to achieve a smaller spot, particularly if the focusing is not diffraction-limited. Such difficulty is a likely scenario, for example, if the 6~eV generation process involves several cascaded steps of non-collinear frequency mixing or if the beam shape of the source is not strictly Gaussian.

\subsubsection{Wave mixing in gases: \texorpdfstring{$\chi^{(3)}$}{X(3) } nonlinear processes} 
To push beyond the transparency limit of solids, one can employ nonlinear frequency conversion via the third-order nonlinear polarization, $\chi^{(3)}$ in gases. Photons in the 9-11~eV range can be generated by pumping a gas cell with the second harmonic of a Ti:sapphire \cite{Cilento2016}, or the third harmonic of a Yb-doped laser source \cite{Lee2020, Peli2020, Kawaguchi2023,Berntsen2011, He2016}. The phase-matching of the THG process in gases is sensitive to the focusing geometry, gas pressure, and the frequencies of the fundamental and third harmonic relative to atomic transitions \cite{Cilento2016, He2016}. In particular, THG seeded by the 2H of the Ti:sapphire FF is impossible to phase match in the tight-focusing geometry due to the normal dispersion of gas (Xe) near 9.3~eV (133~nm). Here, the 3H can only be produced from less efficient six-wave mixing ($4\omega-\omega$) as opposed to the desired four-wave mixing ($2\omega+\omega)$ \cite{Cilento2016}. THG seeded by the 3H of the FF of the Yb-doped laser is much more efficient, as Xe has anomalous dispersion at 10.8~eV (115~nm). Furthermore, for ultrafast TR-ARPES systems, the dispersion of the gas and windows at the relevant photon energies must be accounted for if short pulses are desired.

\begin{figure}[t!]
    \centering
    \includegraphics{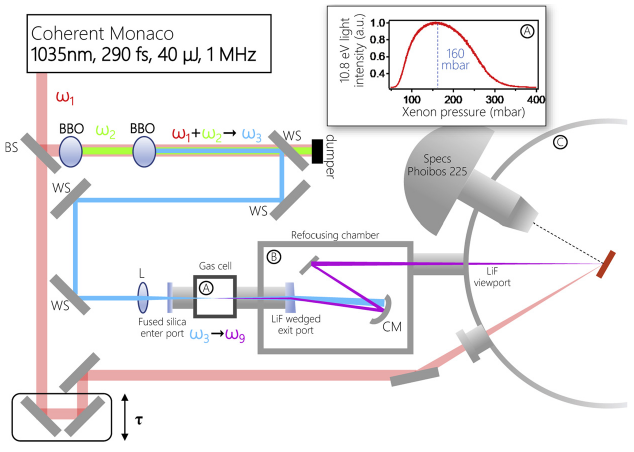}
    \caption{\textbf{Experimental setup of UV-driven third-harmonic generation in gases.} Figure adapted from Peli \textit{et al.} \cite{Peli2020}. BS: beam splitter, WS: wavelength separator, L: lens, CM: curved mirror, BBO: $\beta$-BaB$_2$O$_4$. A: gas cell; B: refocusing chamber; C: ARPES chamber. The fundamental frequency (1035~nm) is up-converted to the third harmonic in BBO and used to seed the third harmonic generation (THG) in Xenon.}
    \label{Fig: THG}
\end{figure}
As seen from Fig. \ref{Fig: Probe_survey}, $\chi^{(3)}$ gas-based sources appear to be under-utilized, given their simplicity and potential for high-resolution ARPES and TR-ARPES. The accessible range of k-space is sufficient to access the full Brillouin zone of some heavily-studied materials like the cuprates. Furthermore, compared to HHG-based systems, they appear to have more straightforward beamline requirements in that they can operate in a nitrogen-purged environment, and LiF or MgF$_2$ vacuum windows can be used at the beamline/UHV chamber interface, thus eliminating the need for a costly differential pumping system (see Fig.~\ref{Fig: THG}).

Two recent TR-ARPES systems utilizing THG in gases are driven by Yb:KGW lasers. Lee {\it et al.} use the third harmonic from a 190~fs, 100-250~kHz repetition rate system to pump a THG stage using xenon gas in a hollow core fiber \cite{Lee2020}. The 11-eV beam is isolated with an off-plane grating monochromator and a 0.5~mm LiF window at the exit \cite{Lee2020}. This system achieves a probe energy resolution of 16~meV and a pump-probe time resolution of  250~fs. Peli {\it et al.} present a similar Yb:KGW laser system with 290~fs pulses at 4~MHz repetition rate. Here, THG is performed in a gas cell, and the system achieves an energy resolution of 26~meV with a time resolution of 700~fs \cite{Peli2020}. Notable differences between these two systems include: (i) the monochromator in Lee \textit{et al.} allows filtering of the spectrum to improve the energy resolution \cite{Lee2020}, and (ii) dispersion in the material (nitrogen gas, LiF window, and LiF prism device) may play a role in the time resolution reported in Peli \textit{et al}. These systems demonstrate that THG sources are useful for TR-ARPES \cite{Baldini2023}, and optimization of the pulse parameters will likely reduce the measurement time-bandwidth product through optimization of the VUV pulse characteristics. Lastly, we note the recent development of a commercial system which utilizes a cascaded four-wave mixing approach in hollow-core fiber at 1 MHz. This system has demonstrated a sufficient photon flux at photon energies as high as $\approx$11~eV \cite{Couch2020}; however, to the best of our knowledge, it has yet to be utilized as a probe for TR-ARPES studies.

\subsubsection{High-order harmonic generation}
\label{Sec: HHG}
\begin{figure}[t!]
    \centering
    \includegraphics{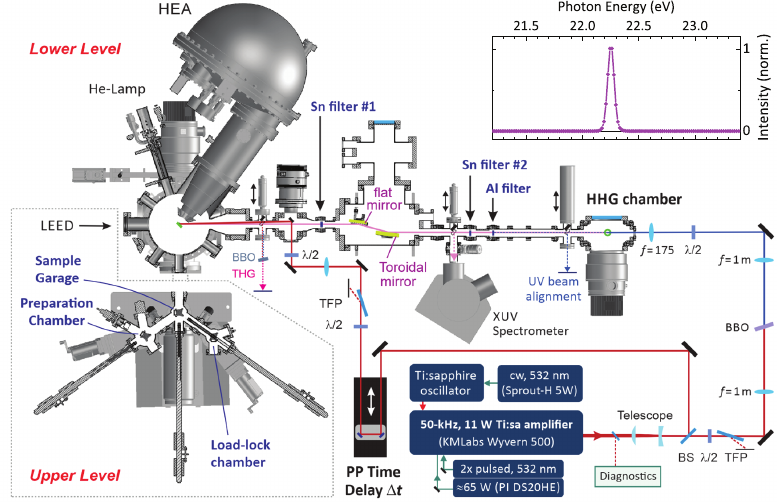}
    \caption{\textbf{Experimental setup of an XUV source.} Figure adapted from Buss \textit{et al.} \cite{Buss2019}. The XUV beamline with toroidal focusing and UHV ARPES endstation is shown. The XUV pulse spectrum measured by the spectrometer is inset. Chambers for sample loading, garage storage, and surface preparation are attached to the upper level of the main chamber. At the same time (TR)-ARPES is performed on the lower level with XUV and helium lamp sources. TFP: thin film polarizer, $\lambda/2$: half-wave plate, BS: beam splitter, BBO: $\beta$-BaB$_2$O$_4$, THG: third harmonic generation, HEA: hemispherical electron analyzer, LEED: low-energy electron diffraction.}
    \label{Fig: Buss_setup}
\end{figure}

To reach probe photon energies higher than 11~eV, one must drive materials beyond the perturbative limit of nonlinear polarization. At high intensity ($>10^{13}$ W/cm$^2$), a series of harmonics --odd harmonics in gases, due to the inversion symmetry-- appears from a process known as high-order harmonic generation, or HHG. In HHG systems, the high average power, high photon energies (15-100~eV) and presence of a nonlinear, laser-induced plasma create a new set of technical challenges beyond those already addressed for the 6-eV and 11-eV systems. To integrate this source with the photoemission UHV chamber, a differential pumping tube is needed to separate the HHG chamber in low vacuum from the measurement chamber in UHV. Additional focusing optics, a filter and/or monochromator, and other beam diagnostics are needed to monitor the stability of the system during operation. To highlight the added complexity in comparison to 6-eV and THG systems, the experimental setup from Buss \textit{et al.} \cite{Buss2019} is shown in Fig.\,\ref{Fig: Buss_setup}. In this section, we will discuss HHG generation, spectral separation, and beamline components sequentially.

\begin{figure}[t!]
    \centering
    \includegraphics{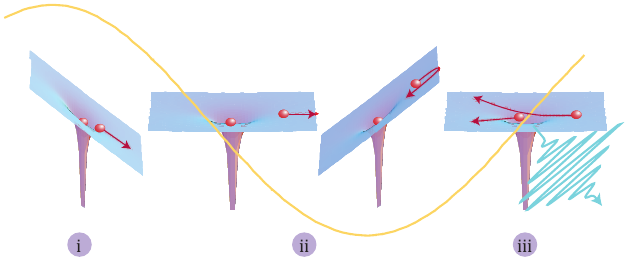}
    \caption{\textbf{The three-step model description of high-harmonic generation in gases.} Figure adapted from Corkum and Krausz \cite{Corkum2007}. (i) An intense femtosecond pulse (near-infrared or visible, yellow) distorts the atomic potential and induces tunnelling ionization of an electron. For ionization in this strong field, the electron motion can be described with Newtonian equations of motion. (ii) Initially, the electron accelerates away from the parent ion; as the field is reversed, the electron is driven back and re-collides during a small fraction of the laser oscillation cycle. (iii) The parent ion sees an attosecond electron pulse, which is converted into an optical attosecond pulse upon recollision (teal).}
    \label{Fig: HHG3step}
\end{figure}
The HHG process is briefly described here following the three-step model of Corkum \cite{Corkum1993} and shown in Fig.\,\ref{Fig: HHG3step} (adapted from Corkum and Krausz \cite{Corkum2007}): (i) the driving field induces distortion in the atomic potential and enables tunnelling ionization of an electron; (ii) the ionized electron trajectory is guided by the laser field, which has some probability of returning to the parent ion with high ponderomotive energy after the first laser period; and (iii) the recollision of the electron with ion generates a burst of high energy radiation. Over many cycles of this process, the spectrum of radiation appears as odd harmonics of the driving field. The harmonics extend to a maximum photon energy given by $h\nu_\text{cutoff}=I_p + 3.17 U_p$. $I_p$ is the ionization potential of the gas. $U_p$ is the pondermotive energy of the field, where $U_p \propto I_L \lambda^2$ for a pulse of wavelength $\lambda$ and peak intensity $I_L$. The fundamentals of the HHG process have been reviewed in the context of macroscopic phase matching in real ensembles of atoms and high average power, high repetition rate sources several times in recent years \cite{Hadrich2016, Heyl2017, Mills2012, Pupeza2021}; we do not review them here, except where it is critical to our understanding of different TR-ARPES sources.

High-harmonic sources in TR-ARPES labs today exist in several categories, shown in Table~\ref{Tab: HHG_classes}. Ti:sapphire systems are usually based on regenerative amplifiers, whereas Yb-doped material lasers are often based on pulse-pickers and single-pass chirped-pulse amplification. These systems cover the repetition rate regime of 10~kHz to 1~MHz. At higher repetition rates, the available pulse energies used to drive the HHG are on the $1-100~\upmu$J scale, which limits the efficiency of harmonic generation. To compensate, most single-pass HHG systems perform the HHG using the second or third harmonic (2H/3H) of the laser fundamental to take advantage of the powerful scaling ($\approx \lambda^{-6}$ \cite{Shiner2009}) of HHG flux with increasing photon energy of the driving field \cite{Guo2022}. Yb-doped amplifier systems on the order of 100~W output power have been utilized in TR-ARPES with repetition rates as high as 1~MHz \cite{Keunecke2020}. In at least one case, an optical parametric chirped pulse amplifier (OPCPA) system based on a Yb-doped laser system is used to produce 1.55~eV pulses, and HHG is performed using the second harmonic of this source at 500~kHz repetition rate \cite{Puppin2019}. To operate at very high repetition rates ($\gtrapprox$10~MHz), HHG can be performed in femtosecond enhancement cavities (fsEC) at the 1.2~eV FF of Yb-doped lasers.

\begin{table}[b!]
\begin{center}
\caption{Summary of laser classes commonly used in lab-based HHG TR-ARPES systems, including repetition rate regime, laser technology and HHG driving field frequency, including fundamental laser frequency (FF), second harmonic (2H) or third harmonic (3H). Ti:sapphire HHG laser systems are typically based on regenerative amplifiers (Regen) and Yb-doped systems leverage the high average power available in single-pass amplifiers (SPA) and optical parametric chirped-pulse amplifiers (OPCPA). The nonlinear medium is usually a gas jet (GJ) or gas-filled capillary (CAP). For repetition rates $>$10~MHz passive resonator cavities combine pulses coherently from the 10-100~W amplifiers to enter the multi-kW regime.}
\label{Tab: HHG_classes}
\begin{tabular}{  c | c | c | c | c  }
 Platform & Rep. Rate & Technology & HHG & Geometry \\ 
 \hline
 Ti:sapphire \cite{Rohde2016,Heber2022} & $\le$10 kHz & mJ Regen & FF & CAP \\  
 Ti:sapphire \cite{Sie2019, Buss2019}  & 50 - 250 kHz & $\upmu$J Regen & 2H & GJ, CAP\\    
 Yb-doped \cite{Cucini2020,Guo2022,Keunecke2020} & 50-1000~kHz & 100 $\upmu$J SPA & 2/3H & GJ\\
 Yb-doped \cite{Puppin2019} & 500 kHz & Yb OPCPA & 3.1~eV & GJ\\
 Yb-doped \cite{Mills2019, Corder2018, Saule2019} & $>$10 MHz & cavity-HHG & FF & GJ\\
\end{tabular}
\end{center}
\end{table}
Each of the platforms listed in Table 1 has distinct technical requirements or challenges that arise due to their use and may present different challenges depending on the expertise of a particular lab. For example, HHG performed with the FF requires spectral separation of more closely spaced harmonics, whereas 2H/3H-HHG produces larger spacing between harmonics, which facilitates better spectral selectivity. Use of 2H/3H-HHG requires more optics capable of handling high-power UV light, which requires careful selection of optics to prevent long-term degradation \cite{Guo2022}. Using an OPCPA for HHG offers the possibility of tunable HHG, which is not usually possible with most Ti:sapphire or Yb-doped laser systems. Possible challenges with these systems involve the very high power pump lasers used in the OPCPA and the high-power UV used for the 2H-HHG from the OPCPA output. Finally, cavity-HHG systems require active stabilization of the cavity and low-loss, low-dispersion cavity mirrors capable of handling kilowatt average power in the near-infrared.

HHG is only one aspect of the light-matter interaction with high-intensity pulses; the driving field also produces ionization, which has a significant impact on the use of HHG as sources of VUV/XUV light \cite{Corkum1993}. Given the highly nonlinear nature of HHG, its efficiency is extremely low. While some fraction of ionized electrons recombine with the parent ion, a much greater fraction remains a plasma. Generally, this plasma should be minimized because it limits the ability to phase-match the harmonics. Several approaches exist to minimize plasma build-up, including the geometry of the gas (capillary versus jet) and the duration of the driving pulse. In practice, plasma build-up is often avoided with shorter driving pulses, but due to the trade-off with the bandwidth of the HHG, this is not desirable for high-energy resolution systems. At very high repetition rates, some plasma may be present between successive pulses, and in the case of cavity-HHG, this persistent plasma also limits cavity enhancement. 

Early work on HHG sources at lower repetition rates (e.g. $<$10 kHz) focused on maximizing HHG efficiency using high pulse energy with weak focusing or guided modes in glass capillaries. The emergence of higher repetition rate sources with much lower pulse energies led to the exploration of HHG efficiency in the tight focusing regime \cite{Heyl2012, Chiang2012, Rothhardt2014, Chiang2015}. Two notable examples with direct links to TR-ARPES experiments include the use of few-$\upmu$J pulses to produce HHG from a commercial Ti:sapphire laser \cite{Heyl2012}, and 5-10~$\upmu$J pulses in a Ti:sapphire cavity-HHG system \cite{Hammond2011}. The former demonstration led the authors to develop a macroscopic theory of the scaling properties of HHG efficiency \cite{Heyl2017}. These scaling laws allow one to understand how to preserve the properties of the HHG process while changing laser pulse energy and focusing parameters (focus size and divergence) or gas parameters (gas interaction length, density, and pressure). A detailed comparison of the relative efficiencies of HHG sources used for TR-ARPES is beyond the scope of this review. Instead, we focus on the technical challenges of bringing a TR-ARPES apparatus into operation. With the continued development of new sources, we will likely see further optimization of XUV generation in different photon energy ranges or pulse duration regimes that take advantage of the knowledge gained through these in-depth studies of the macroscopic HHG process. 

With this examination of existing platforms for HHG in mind, we move to a discussion of the probe beamline, an umbrella term encompassing all the components that deliver the generated harmonics to the sample for photoemission. Generally speaking, it is very challenging to produce all the desired characteristics simultaneously, including high flux throughput, high fidelity separation of harmonics, preservation of ultrashort pulse duration, focus to a small spot on the sample, and arbitrary polarization control. The high average power of present-day TR-ARPES laser systems creates engineering challenges from the perspective of mechanical stability and material damage or degradation, particularly for intense visible and UV-driving lasers. 

To maximize the XUV flux, the high photon energies in probe beamlines require special considerations, including the use of first-surface reflective components for (roughly) the 10-30~eV range and sometimes multi-layer dielectric mirrors for higher photon energies. Even with advanced XUV optics, each reflection results in a significant loss of XUV flux, such that care must be taken to design the beamline with as few optics as possible, yet maintaining the necessary degrees of freedom to achieve the desired alignment and focusing with minimal aberrations. 

A well-known problem in XUV beamlines is the decrease in the beamline efficiency due to the contamination of optical elements by hydrocarbon build-up or other contaminants. Both a high XUV flux and the driving field (UV to IR) can contribute to this process \cite{Keunecke2020}. In cavity-enhanced HHG sources, hydrocarbon contamination is particularly problematic because the loss of mirror reflectivity affects the enhancement of the fundamental field inside the resonator \cite{Mills2019, Corder2018}. The contamination problem is mitigated through a continuous flow of a mixture of oxygen and ozone, produced with inexpensive commercial ozone generators, into the vacuum chamber. In some cases, it may be helpful to introduce the ozone close to mirrors that experience the greatest contamination issues; in other cases, it is sufficient to introduce a small flow into the entire vacuum chamber far away from problematic components. In some situations, it has been useful to supplement the in-situ cleaning process with infrequent external plasma cleaning of select mirrors \cite{Mills2019}, but only after several months of continuous operation.

\begin{figure}[t!]
    \centering
    \includegraphics{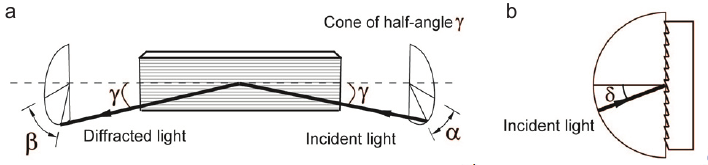}
    \caption{\textbf{The time preserving monochromator (TPM), adapted from Frassetto \textit{et al.} \cite{Frassetto2011}}. \textbf{(a)} The geometry of the off-plane mount. Incident light arrives along a cone of half-angle $\gamma$ at an azimuthal angle $\alpha$. The first diffraction order leaves on the same cone at an azimuthal angle $\beta$. \textbf{(b)} View of the TMP from the side highlighting the blaze angle. The efficiency is maximized when $\alpha=\beta=\delta$.}
    \label{Fig: TPM}
\end{figure}

Next, we consider the spectral separation and isolation of the harmonics. The driving field and the harmonics propagate collinearly from the interaction region with the gas. A number of strategies have been used to separate the driving field from the harmonics, including thin film metallic filters \cite{Sie2019, Buss2019}, spatial filtering based on the difference in divergence of the laser and harmonics \citep{Cucini2020, Guo2022}, or using grazing incidence anti-reflection plates \cite{Keunecke2020}. With respect to the spectral separation of the harmonics themselves, multi-layer mirrors can be designed to reflect specific harmonics \cite{Guo2022} selectively. However, the extinction of adjacent harmonics is not perfect, which can lead to a spurious intensity above $E_F$ due to photoemission from a higher-order harmonic. Grating monochromators are typically not ideal for ultrafast pulses due to the wave-front tilt introduced by the grating \cite{Bor1993}. In short, the wave-front tilt is due to the total optical path difference $\Delta \text{OP} =\lambda N$ in the diffracted beam, where $\lambda$ is the wavelength of the light and $N$ is the number of illuminated grooves. Monochromators with multiple gratings and 1~fs level time compensation can be designed at the expense of photon flux, as each reflection is significantly lossy at these high photon energies \cite{Poletto2006}.  

In this respect, an extremely valuable development has been that of a high efficiency, low groove density single-grating design in the off-plane mount with a temporal response of a few tens of fs, shown in Fig.\,\ref{Fig: TPM} \cite{Frassetto2011, Poletto2012}. In the off-plane mount, the grating grooves are parallel to the plane of incidence of the light. In this geometry, the projection of the beam spot on the grating due to the grazing incidence geometry does not illuminate any additional grooves, and the pulse-front tilt is reduced by a factor of $1/\sin(\gamma)$ -- around 5 to 10. The grating equation is
\begin{equation}
    \sin(\gamma)[\sin(\alpha)+\sin(\beta)]=\lambda_n\sigma,
\end{equation}
where $\alpha$ and $\beta$ are the azimuth angles of the incident light and the first-order diffracted $n$th harmonic ($\lambda_n$), respectively, $\gamma$ is the grazing angle of the beam incident on the grating, and $\sigma$ is the density of grooves. In this way, the grating can be rotated such that $\alpha=\beta$ to select the $n$th harmonic. Since the efficiency of the grating changes with $\beta$, the gratings can be exchanged to adjust properties such as focusing characteristics and the maximum efficiency wavelength, making this option an excellent choice for widely-tunable XUV beamlines. 

Lastly, we discuss design considerations with respect to the tuning of the probe parameters, such as photon energy, spot size, and polarization. One of the advantages of HHG is that it produces a spectrum of harmonics in the 10 to 100~eV range, all of which --in principle-- can be used for photoemission. This added flexibility is advantageous for leveraging the photon energy-dependent photoemission matrix elements, which can greatly enhance the intensity of spectral features. In practice, however, it is impossible to optimize all harmonics simultaneously, and gratings are designed with high efficiency for only a select photon range. Therefore, achieving tunability over the full spectrum is typically not possible without dedicated design and user effort, while tunability over a range of 10 to 15~eV is much more achievable over the course of a single experiment. 

With respect to spot size, focusing XUV pulses to $10~\upmu$m should be straightforward due to the short wavelength. However, the numerous beamline requirements described previously greatly increase the technical challenge. The spot size can be limited by the surface quality of large, grazing-incident components or aberrations that arise when easily manufactured mirror forms are chosen over optimal theoretical design (e.g. toroidal vs. elliptical mirrors). The desired spot size produced at the sample should be a primary consideration when designing the beamline and the interface to the UHV chamber. As we discussed in Sec.~\ref{Sec: Technical}, a decrease in the spot size of the probe leads to increased SC distortions. With presently used TR-ARPES lasers systems, SC is observed at nearly any repetition rate if the beams are small enough. Two notable examples are observed at 200~kHz with a 100~$\upmu$m spot size for individual harmonics on gold \cite{Cucini2020}, and at 18~MHz with a 10~$\upmu$m spot on tungsten \cite{Saule2019} when the full HHG spectrum is focused onto the sample.

Finally, Sec.\,\ref{Sec: ARPES intensity} discussed the role of polarization in the investigation of material properties in ARPES experiments. In HHG, the polarization of the XUV is linear and parallel with the driving field. It is possible to generate elliptical or circular polarization for dichroism experiments, both through the generation process or by using subsequent optics in the beamline. Still, both approaches are challenging and can come with a considerable loss of flux compared to the linearly polarized sources. Currently, the development of XUV sources with tunable polarization for TR-ARPES is lacking. However, given significant motivation from the condensed matter perspective, they will likely emerge soon. We briefly discuss these prospects in Sec.~\ref{Sec: ExpDrivenParams}.

\subsection{Pump sources}
\label{Sec: Pump sources}
In this section, we discuss methods used to generate the pump beam for TR-ARPES. As in other pump-probe spectroscopies, this pulse perturbs the material system from equilibrium and relaxation processes return the system to the same equilibrium state or a meta-stable state. Here, we consider situations where the pump has photon energy below the work function of the material. The pump is usually derived from the same laser oscillator as the probe to minimize pump-probe timing jitter, but it is also possible to synchronize two separate lasers to achieve a stable pump-probe delay \cite{Guo2022}. The FF of the laser and its second harmonic are often the first photon energies used with a new TR-ARPES system; however, tuning the pump photon energy can enable the study of much new and exciting physics. Nonlinear frequency conversion techniques allow the photon energy of the pump to be tuned over nearly the entire range from the THz to the visible, often with commercially available tools. The pump pulse duration is another means of accessing different non-equilibrium observables, as we discuss briefly in Sec.~\ref{Sec: ExpDrivenParams}.  

\begin{figure}[t!]
    \centering
    \includegraphics{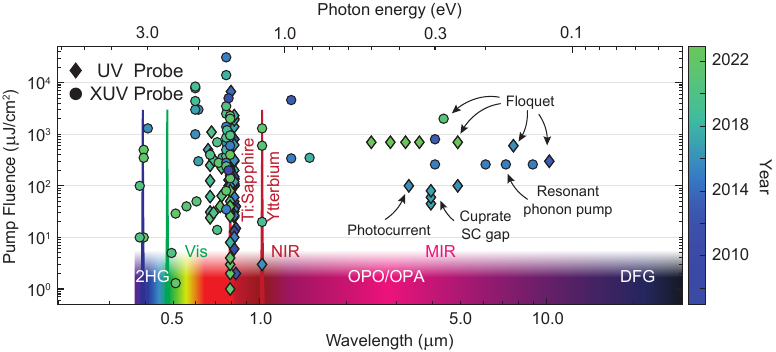}
    \caption{\textbf{Overview of pump sources used in TR-ARPES experiments.} The pump fluence is plotted as a function of photon energy. Many studies have been performed with the pump at the Ti:sapphire fundamental frequency (FF) (780-840~nm) at all fluences. More recently, the OPO/OPA has extended the tunability of the pump into the mid-infrared. This regime has been used to demonstrate Floquet physics and the resonant pumping of bosonic modes. The second harmonic of the FF and OPA outputs and the 2H-driven OPA reaches most of the visible spectrum. Excitons and carrier dynamics in semiconductors and molecules are accessible here.}
    \label{Fig: Pump survey}
\end{figure}
\begin{figure}[t!]
    \centering
    \includegraphics{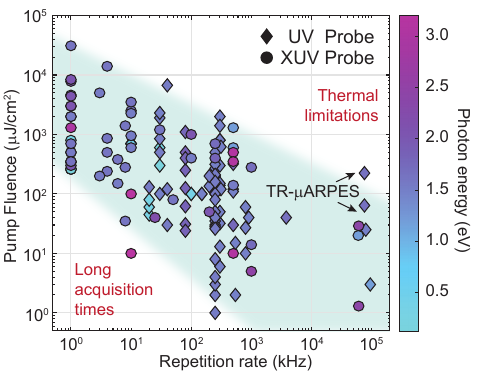}
    \caption{\textbf{Survey of pump parameters: Fluence and repetition rate.} Pump fluence as a function repetition rate of TR-ARPES studies. The available fluence is loosely constrained to the repetition rate, with $\approx$ 1-2 orders of magnitude of tunability. The low-fluence, low-repetition rate regime is limited by intractably long acquisition times. The high-fluence, high-repetition-rate regime is limited by thermal damage to the sample. References in this figure can be found in Tab.\,\ref{Tab: pump_survey}.}
    \label{Fig: Pump param}
\end{figure}

To survey the pump sources for TR-ARPES, we examine pump pulse parameters sourced from the ``scientific output" of TR-ARPES publications. These are a better reflection of the state of TR-ARPES than ``technical" reports, which represent the state of nonlinear optics research. The references in this survey (shown in Fig.\,\ref{Fig: Pump survey} and Fig.\,\ref{Fig: Pump param}) are collected in Tab.\,\ref{Tab: pump_survey}. We show the two most varied parameters (pump fluence and photon energy) in Fig.\,\ref{Fig: Pump survey}. The colour of the marker indicates the year in which the paper was published, and the shape of the marker denotes the photon energy of the probe. What is immediately striking is the sheer number of studies that have been done with a pump at the Ti:sapphire fundamental of $\approx 1.55$~eV. These studies span the entire fluence range and the modern development of TR-ARPES. Part of this is due to the proliferation of Ti:sapphire lasers for TR-ARPES and other ultrafast spectroscopic techniques. The photon energy is larger than the bandgap of many semiconductors, and the fluence, polarization, pulse duration, and spot size are easily tunable with conventional optics. In contrast, results from new Yb-doped sources are sparse. Unlike Ti:sapphire sources, the wavelength of Yb-doped sources (1030~nm or 1.2~eV) is below many semiconductor band gaps, limiting the use of the fundamental wavelength as the pump in studying carrier dynamics and excitons in many materials. 

To tackle this lack of flexibility, many Yb-doped (and even Ti:sapphire) sources have integrated an optical parametric oscillator (OPO) and/or amplifier (OPA) to tune the frequency in the near and mid-infrared \cite{Bovensiepen2007}. There are many attractive opportunities for pumping solid-state materials in this frequency range. Here, the photon energy can be tuned to resonate with low-frequency collective excitations such as phonons and magnons; or be used to probe the low energy dynamics of correlated electronic phases, such as unconventional superconductors and charge/spin density waves \cite{Basov2011}. Like their high-frequency counterparts, these pulses can also be used to control material properties in an ultrafast manner through driving nonlinear phononics \cite{Forst2011, Henstridge2022} or inducing hybrid light-matter states \cite{Wang2013}. Difference frequency generation (DFG) seeded by the signal and idler outputs of an OPA allows for efficient generation of pulses in the multi-THz frequency range ($>10$~THz) \cite{Gierz2015}. When utilized as the pump in TR-ARPES studies, these sources have enabled the investigation of Floquet-Bloch states in topological insulators \cite{Wang2013, Mahmood2016}, graphene \cite{Aeschlimann2021}, and black phosphorus \cite{Zhou2023}. To date, most mid-infrared TR-ARPES experiments have been performed at relatively high fluence, largely due to the low repetition rate of the sources. Subsequent frequency up-conversion of the OPA outputs into the visible and UV are high enough in photon energy to overcome the bandgap of most semiconductors. While unwanted multi-photon photoemission events somewhat curtail the fluence of these high-photon energy pumps, many recent results have shown their capability in studying semiconductors \cite{Lee2021}, dynamic bandgap renormalization \cite{Ulstrup2016, Chen2018, Roth2019, Hedayat2021} and resonant excitation of excitonic states \cite{Madeo2020, Wallauer2021, Schmitt2022, Kunin2023, Lin2022}.

The trade-off between ECR and pump power was highlighted in Sec.\,\ref{Sec: Technical}. The impact of the trade-off is clearly shown in the survey of pump parameters shown in Fig.\,\ref{Fig: Pump param}. At low repetition rates where ECR is low, the acquisition time needed to discern the pump-induced signal at low fluence is prohibitively long; hence, these experiments typically have $F>100~\upmu$J/cm$^2$. At high repetition rates, the sample is prone to average power-induced thermal damage, and experiments are generally performed with $F<100~\upmu$J/cm$^2$. Higher fluence at high repetition rates can be achieved, however, if the spot size of the pump and probe can be reduced to the order of $<10~\upmu$m (see TR-$\upmu$ARPES points in Fig.\,\ref{Fig: Pump param}) or if the sample is resistant to thermal damage. We also see that below $10$~kHz repetition rate, experiments are performed exclusively using XUV probes. These come from the pre-2015 Ti:sapphire-based HHG sources in Fig.\,\ref{Fig: Probe_param}b. Later HHG sources at 100~kHz to 10~MHz expanded the usable fluence to the 1-10~$\upmu$J/cm$^2$ regime.

\subsubsection{Nonlinear Frequency Conversion Based on Optical Parametric Processes}
\begin{figure}[t!]
    \centering
    \includegraphics{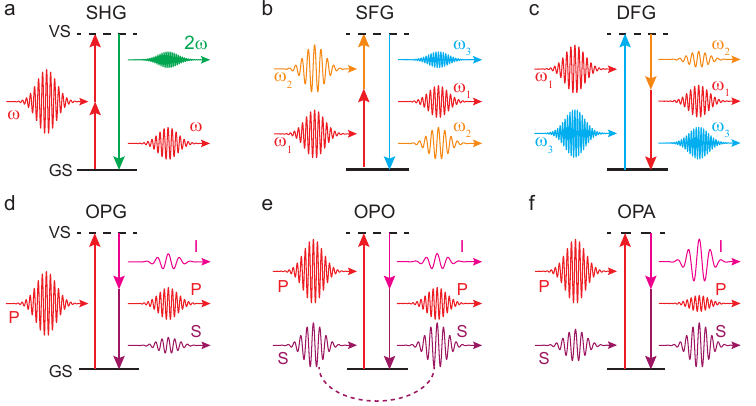}
    \caption{\textbf{Frequency tuning the pump with optical parametric processes.} Ground/virtual states as denoted as G/VS. \textbf{(a)} Second harmonic generation (SHG): two photons of frequency $\omega$ add to produce $2\omega$. \textbf{(b)} Sum frequency generation (SFG): two photons of frequencies $\omega_1$ and $\omega_2$ add to produce $\omega_3=\omega_2+\omega_1$. \textbf{(c)} Difference frequency generation (DFG): two photons of frequency $\omega_3$ and $\omega_2$ produce $\omega_1=\omega_3-\omega_2$. \textbf{(d)} Optical parametric generation (OPG): a photon ($\omega_\text{pump}$) spontaneously converts into two lower energy photons ($\omega_\text{signal}$ and $\omega_\text{idler}$) inside a nonlinear medium. \textbf{(e)} Optical parametric oscillator (OPO): The nonlinear crystal is inside a cavity resonant with the signal wavelength. The signal produced in the nonlinear medium recirculates to seed the optical parametric process. OPOs are typically used to amplify the signal strength up to the limit imposed by losses in the cavity. \textbf{(f)} Optical parametric amplification (OPA): Similar to a DFG process with a weaker seed (the signal). The OPA is used to further amplify the signal and idler pulse energy.}
    \label{Fig: OPP}
\end{figure}
The two primary laser platforms we are considering here (i.e. Ti:sapphire and Yb-doped media) both operate in the near-infrared, and desirable pump frequencies exist above and below the laser fundamental. Frequency-tuning is usually performed with optical parametric processes where energy is transferred from one or more input waves to other frequencies via the second-order nonlinear polarization, $\chi^{(2)}$, of a material. A general overview of optical parametric processes is given in Fig.\,\ref{Fig: OPP}, and a recent review details these processes \cite{Vodopyanov2020}. Second harmonic generation (SHG) is an example of sum frequency generation (SFG), where two photons of the same frequency add to produce a wave of twice the fundamental frequency. Generalized to non-degenerate frequencies, wave mixing between three fields can be set up to perform SFG, DFG or amplification of an input wave via optical parametric amplification (OPA). The OPA is particularly important for its application to tunable near-infrared and mid-infrared radiation. Specifically, the OPA refers to the process whereby a strong `pump' wave transfers its energy to a weaker `signal' wave in a nonlinear medium. Through this process, a third wave or `idler' is also produced such that $\omega_\text{signal}+\omega_\text{idler} = \omega_\text{pump}$. However unfortunate, the use of `pump' to describe both the OPA pump and the TR-ARPES pump is unavoidable.

The specific requirements for a tunable source based on parametric processes largely depend on the pulse energy of the pump. With a suitably high pulse energy (approximately in the $\upmu$J range), a white light continuum can be generated from the laser using a nonlinear medium such as yttrium-aluminum-garnet (YAG) or sapphire and used as the seed for the OPA. Here the phase-matching condition in the crystal is adjusted to produce the desired signal and idler wavelengths. In the high repetition rate (low pulse energy) regime, the OPA seed needs to be generated using optical parametric generation (OPG) or an optical parametric oscillator (OPO). Both OPGs and OPOs build up the signal and idler power from spontaneous emission. They thus can be susceptible to issues with noise properties or parasitic effects depending on the design. OPGs (Fig.\,\ref{Fig: OPP}d) do not have special path length requirements and can be very compact, while OPOs (Fig.\,\ref{Fig: OPP}e) need a resonant cavity operating at the signal wavelength and a round trip time equal to the pump source, which can be technically challenging to stabilize. Still, for higher repetition rates --particularly the tens of MHz rates of mode-locked oscillators-- the OPO resonator length is manageable and can incorporate optical fiber in the cavity to realize a compact and robust design \cite{Steinle2016}. The relatively modest output power of the signal and idler in the OPO/OPG stages can be subsequently amplified with an OPA.

\subsubsection{Frequency tuning from visible to mid-IR}
Optical parametric processes have been used for decades and are well documented in the literature \cite{Vodopyanov2020, Gaida2018, Tian2021}. Yet, new nonlinear materials technology has enabled scaling to higher average power and pushed nonlinear frequency conversion further into the infrared and THz. With the high average power lasers presently employed, OPA and OPO sources are available commercially to produce TR-ARPES pump sources that span the visible to infrared range using one laser. In selecting a particular approach for a new TR-ARPES system, two key considerations are the frequency of the fundamental driving laser (generally centred at 1.55 or 1.19 eV) and the repetition rate because it determines the pulse energy regime, as described in the previous section. 

The laser output can pump an OPA directly to generate near-infrared to mid-infrared light. Using the FF of a Yb-doped laser system, a signal and idler can be produced to span (roughly) the 1.4-4~$\upmu$m (0.3- 0.9~eV) range, and using a Ti:sapphire roughly 650~nm-2.5~$\upmu$m (0.5-1.9~eV) can be generated. To extend tuning further into the IR, DFG between the signal and idler outputs of the OPA can be used to produce light well beyond the 10~$\upmu$m (0.12~eV) spectral range. Recent advances in high-average power laser systems and materials for IR nonlinear optics have created opportunities to produce long-wavelength IR in novel ways. For example, the nonlinear properties of new materials have enabled direct IR generation of IR radiation in the 2.5-10~$\upmu$m range with a two-stage OPA, where the first stage is pumped with the second harmonic of the output of a Yb-doped laser system to generate pulses in the 1.15-1.70~$\upmu$m range \cite{Villa2021}. These pulses act as the signal in a second OPA pumped with the FF of the Yb-doped laser system at 1.03~$\upmu$m. The resulting idler wavelength is tunable from 2.5-10~$\upmu$m.

If frequencies higher than the laser fundamental are desired, an OPA can be pumped with the second harmonic of the laser output to produce visible and near-infrared light. For a Yb-doped laser, this produces roughly a tuning range of 650~nm-900~nm (1.38-1.9~eV) for the signal and 1200-2500~nm (0.5-1.0~eV) for the idler. These numbers provide a sense of what is possible, but are not a definitive range. For all of these configurations, the conversion efficiency and the tuning range are sensitive to the pulse duration, pulse energies and the nonlinear crystals used for the frequency conversion, which must all be considered carefully in the design of the OPA \cite{Manzoni2016}.

As a practical matter, a TR-ARPES pump beamline consists of numerous optical elements to control beam shape, power, polarization and beam steering and may include optics inside the vacuum. Antireflection-coated optics are likely needed to maintain the high overall efficiency of the beamline, and dispersion of optical elements must be considered for extremely short pulses. This can be a very challenging task, particularly at the extreme end of the range of system parameters, such as the edges of tuning ranges of the signal, idler, or DFG, and generally for systems at very high repetition rates. The rapid growth in IR generation capabilities has also come with increased availability of IR-optimized optical coatings and components. Despite these challenges, many opportunities exist to generate new scientific results with the expanded pump capabilities in today's laser systems.   

\subsubsection{THz-ARPES}
\label{Sec: THz}
While Fig.\,\ref{Fig: Pump survey} surveys results in the visible and mid-infrared regime, both single and few-cycle THz has been proposed for TR-ARPES experiments as well \cite{Schwarz2020}. To date, the integration of THz pulses with TR-ARPES (THz-ARPES) has been limited. The first demonstration of THz-ARPES comes from the work of Reimann \textit{et al.} \cite{Reimann2018}. Here, a Ti:sapphire amplifier at 3~kHz seeds both the intense THz pulse and a 6-eV probe. The single-cycle THz pulses were generated by tilted-pulse front (TPF) optical rectification in cryogenically cooled lithium niobate (LiNbO$_3$) crystal. THz pulses with peak electric fields in the range of 2.4~kVcm$^{-1}$ to 25~kVcm$^{-1}$ were used in the experiment. 
\begin{figure}[t!]
    \centering
    \includegraphics{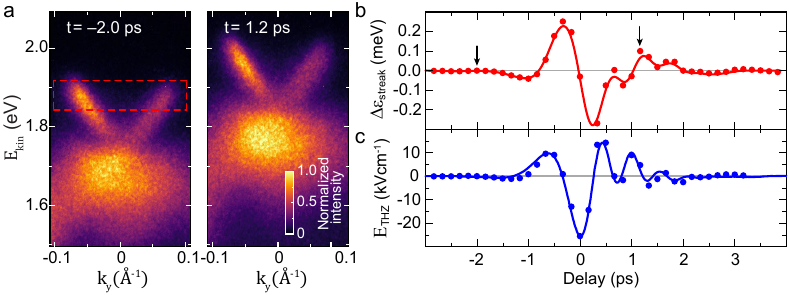}
    \caption{\textbf{Photoelectron streaking by a p-polarized THz pulse.} Figure adapted from Reimann \textit{et al.} \cite{Reimann2018}. \textbf{(a)} Photoemission spectra before the THz pump arrival (t=$-2.0$~ps) and at 1.2~ps after the arrival THz field maximum. The dashed red rectangle indicates the region used to evaluate the time-dependent energy shift. \textbf{(b)} Energy streaking as a function of the pump–probe delay. Red circles are extracted from the spectra; the solid curve is a simulation. Arrows indicate the delays of the spectra in \textbf{(a)}. \textbf{(c)}, Reconstructed electric field amplitude of the standing wave in front of the surface (blue circles). The blue solid curve shows the scaled and time-shifted electric field measured externally by electro-optic sampling.}
    \label{Fig: THzARPES}
\end{figure}

In Reimann \textit{et al.} \cite{Reimann2018}, the s-polarized THz pulse drives a photocurrent in the topological surface state of Bi$_2$Se$_3$, which is observed in the photoemission spectra as a delay-dependent displacement of the Fermi surface. However, the photoemitted electrons are also streaked by the THz pulse in the vacuum, which leads to a shift of the spectra in energy and momentum for p- and s-polarized THz pulses, respectively. Fig.\,\ref{Fig: THzARPES}, adapted from Reimann \textit{et al.}, showcases the effect of energy streaking induced by p-polarized THz pulses. The shift of the spectra before THz arrival and at 1.2~ps after the peak electric field is shown in Fig.\,\ref{Fig: THzARPES}a. By measuring the shift as a function of pump-probe delay (shown in Fig.\,\ref{Fig: THzARPES}b), one can reconstruct the THz field at the sample surface (Fig.\,\ref{Fig: THzARPES}c). Here we see that for the 25~kVcm$^{-1}$ p-polarized peak field, the energy streaking is as large as 200~meV. 

\begin{figure}[t!]
    \centering
    \includegraphics{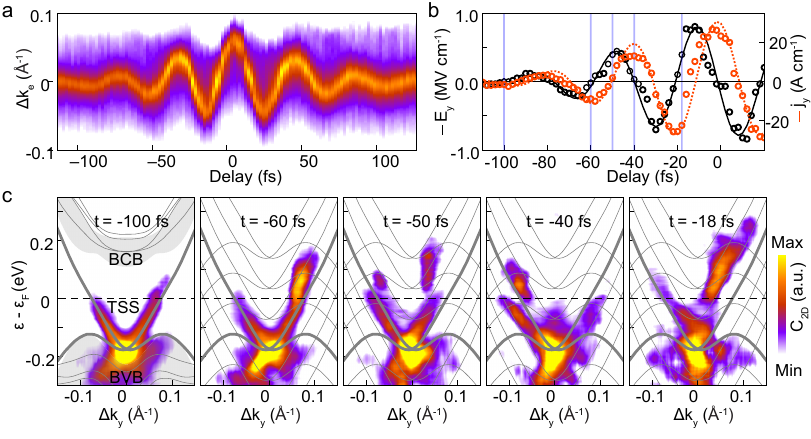}
    \caption{\textbf{Photoelectron streaking by a s-polarized THz pulse and subcycle observation of Floquet-Bloch states.} Figure adapted from Ito \textit{et al.} \cite{Ito2023}. \textbf{(a)}  Momentum-streaking as a function of the pump–probe delay, measured from the shift of the topological surface state (TSS). Note that the intensity here is the second derivative (curvature) of the ARPES intensity. \textbf{(b)} The black markers indicate the reconstructed electric field amplitude of the standing wave in front of the surface, while the red markers indicate the current induced by the THz pulse. The black solid line is a fit of the electric field with a peak amplitude of 0.8~MVcm$^{-1}$, and the dotted red line is the current induced by the black wave according to the semiclassical Boltzmann model. \textbf{(c)} Curvature-filtered ARPES maps as a function of pump-probe delay. The momentum-streaking in panel a is corrected, and the DFT calculation of Bi$_2$Te$_3$ is superimposed in the leftmost panel. The electronic distributions are initially accelerated along the ground-state band dispersion. Subsequently, they split into multiple branches that follow the approximate TSS band structure shifted by integer multiples of the driving frequency (thin grey curves).}
    \label{Fig: ItoThz}
\end{figure}

In a more recent study, a multi-cycle s-polarized MIR pulse (0.1~eV, 25~THz) with an incident field strength of 7~MVcm$^{-1}$ is used to pump the topological insulator Bi$_2$Te$_3$ \cite{Ito2023}. While the pump is generated by difference frequency mixing of two OPA outputs at 1.2 and 1.3~$\upmu$m, we include it here as the probe pulse is shorter than one optical cycle of the pump. The induced momentum-streaking ($\Delta k_e$) induced by the s-polarized pump is shown in Fig.\,\ref{Fig: ItoThz}a, with an amplitude as large as $0.07$\AA$^{-1}$ at the peak field strength. Although the authors corrected for this momentum shift, a smaller shift in energy leads to a distortion of the measured electronic structure. At the peak intensity, this leads to a change in the Fermi velocity of $\approx 20\%$. While the distortion is not particularly deleterious to this study, it may imply problems for photoelectron detection and spectra correction at the larger field amplitudes desired for resonantly driving bosonic modes and nonlinear phononics. 

The reconstructed electric field and current density (measured from the asymmetry of electron population in the topological surface state) are shown in Fig.\,\ref{Fig: ItoThz}b as black and red markers, respectively. By using a sub-cycle probe pulse, the authors can track the generation of Floquet-Bloch sidebands, the intraband acceleration of electrons (Fig.\,\ref{Fig: ItoThz}c), as well as the eventual scattering and dephasing of the Floquet-Bloch states caused by scattering with bulk state electrons. This result constitutes a remarkable advancement in THz-ARPES, requiring carrier-envelope phase stable THz pulses and probe pulses with a sub-cycle time resolution of 17~fs.

There is a wide variety of THz emitters, and we refer the reader to some excellent reviews \cite{Dhillon2017, Bull2021}. The TPF optical rectification in LiNbO$_3$ employed in Reimann \textit{et al.} is a well-established platform for generating intense single-cycle THz pulses. These are compatible with the FF of Ti:sapphire and Yb-doped lasers; however, the pulse energies required to generate the THz pulses are in the $\upmu$J to mJ regime. Combined with the high pulse energy requirement of the probe branch, the repetition rate of the source is severely limited. As with the pump sources above, the suitability of a particular THz emitter depends on many parameters, such as its repetition rate, tunability, bandwidth, pulse duration, and achievable electric field strengths. The integration of a THz emitter with the ARPES UHV system, with its associated optics and focusing constraints, must also be considered. As demonstrations of THz-ARPES are limited, more exploration is needed before the compatibility of these emitters for TR-ARPES can be determined.

\section{Future directions}
\label{Sec: outlook}
\subsection{Science-driven source parameters}
\label{Sec: ExpDrivenParams}
Thus far, we have surveyed the literature on TR-ARPES sources and discussed general trends that emerge due to fundamental and technical aspects of photoemission. Although some of these parameters --such as the pump photon energy and polarization-- are deliberately tuned for certain experiments, often parameters such as repetition rate, fluence, and time-bandwidth product are determined by the availability of sources. We can see examples of this bias in the limited repetition rates ($<10$~KHz) of early XUV probes (Fig.\,\ref{Fig: Probe_param}b) and the huge number of publications using the Ti:sapphire fundamental at 1.55~eV (Fig.\,\ref{Fig: Pump survey}). As ultrafast laser technology continues to evolve and expands the source capabilities for TR-ARPES, the sparse areas of Fig.\,\ref{Fig: Pump survey} and Fig.\,\ref{Fig: Pump param} will be sampled. In this section, we discuss some source parameters that will be important for the next generation of TR-ARPES experiments.
\begin{figure}[t!]
    \centering
    \includegraphics{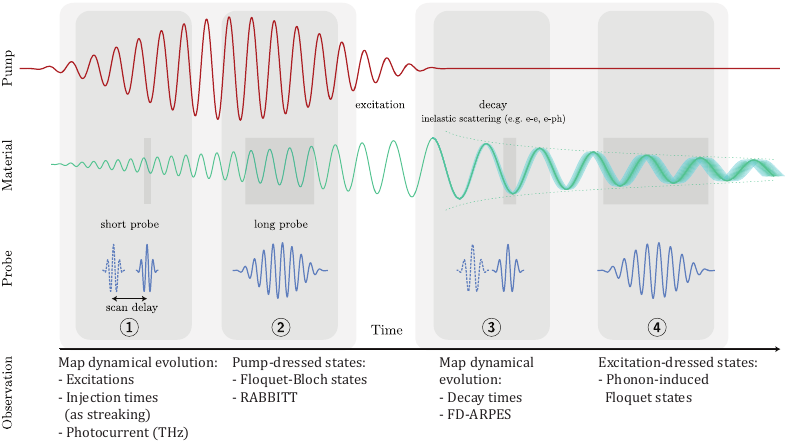}
    \caption{\textbf{TR-ARPES in different pump and probe pulse duration regimes.} Figure adapted from De Giovannini \textit{et al.} \cite{Degiovannini2022}. The pump (red) creates an excitation in the material (green) which is probed using photoemission (induced by the blue pulse). \textcircled{1} and \textcircled{2} distinguish the short and long pulse regimes when pump and probe are overlapped, while \textcircled{3} and \textcircled{4} distinguish the short and long pulse regimes during the material's relaxation process. The short pulse regime is defined as that where the probe duration is less than a single cycle of the pump carrier, whereas the long pulse regime is defined as that where the probe duration averages over many cycles of the pump carrier.}
    \label{Fig: Degiovannini}
\end{figure}

One of these parameters is the probe and pump pulse duration, which --in addition to defining the time resolution-- allows one to access fundamentally different non-equilibrium observables. Following De Giovannini \textit{et al.} \cite{Degiovannini2022}, the discussion is divided into the short and long pulse regimes both during and after excitation (see Fig.\,\ref{Fig: Degiovannini}). In the limit of the short pulse regime, the probe pulse duration is shorter than a single cycle of the pump carrier (\textcircled{1} in Fig. \ref{Fig: Degiovannini}). Examples include single-cycle terahertz pulses, as discussed in Sec.\,\ref{Sec: THz}, and at the short probe wavelengths such as in attosecond streaking chronoscopy \cite{Ossiander2018, Cavalieri2007}. In the short pulse regime, one can accurately map the dynamical evolution of the photoexcitation (or photoemission) process and the rise time of the population of electronic states. After excitation, the short pulse regime refers to the scenario in which the probe pulse is much shorter than a single cycle of the excited bosonic modes in the system (\textcircled{3} in Fig. \ref{Fig: Degiovannini}). Here, one can discern the response of the electronic structure to excited bosonic modes in the system, such as phonons \cite{Sobota2014, Gerber2017}. By taking a Fourier transform of the oscillation and filtering in the frequency domain (also called frequency domain or FD-ARPES), one can directly access the momentum and band-resolved electron-phonon coupling constant \cite{DeGiovannini2020, Hein2020}.

In the limit of the long pulse regime, one can observe light-matter hybrid states, where the electronic Hamiltonian is dressed by the periodic potential of the pump pulse (\textcircled{2} in Fig. \ref{Fig: Degiovannini}). Using a probe pulse that averages over many cycles of the pump carrier, one can study Floquet-Bloch states -- which are the photon-dressed electronic states of the bulk or surface state of the material \cite{Wang2013, Mahmood2016, Zhou2023}. Floquet-Bloch states need not be induced by light specifically; periodic potentials generated by the excited bosonic modes, such as phonons, may also be described by the same formalism \cite{Hubener2018} (\textcircled{4} in Fig. \ref{Fig: Degiovannini}). The pump photons can also dress free electron states in vacuum (laser-assisted photoemission or LAPE). Using LAPE states, one can access photoemission timing information using RABBITT (reconstruction of attosecond beating by interference of two-photon transitions) \cite{Tao2016, Gebauer2019}.

In addition to non-equilibrium observables, parameters --such as probe polarization-- can be varied to elucidate equilibrium observables. As discussed in Sec.\,\ref{Sec: ARPES intensity}, studies using linear and circular dichroism are well established in equilibrium ARPES. In the ultrafast regime, there have been relatively few experimental results, all of which use 6-eV probe pulses \cite{Boschini2020, Zhang2021, Zhang2022b}. This lack is partly due to the availability of ultrafast sources in the XUV with tunable polarization. Since HHG relies on the driving-field-induced re-collision of an ionized electron and the parent ion, ellipticity in the driving field can cause the electron to ``miss" the parent ion, greatly diminishing the HHG yield \cite{Budil1993}. Efforts to address this limitation have yielded several solutions. In the generation process, circularly polarized harmonics have been produced using bi-chromatic driving fields \cite{Fleischer2014, Kfir2015}, non-collinear counter-rotating circularly polarized driving fields \cite{Hickstein2015}. In the beamline, linearly polarized XUV can be converted to circular using transmissive \cite{Schmidt2015} or reflective optics \cite{Vodungbo2011}, although the efficiency of these optics must also be considered.

\subsection{TR-\texorpdfstring{$\upmu$}{micro}ARPES and STARPES}
\label{Sec: smallspot}
In addition to increasing the flexibility and capabilities of TR-ARPES, efforts to combine TR-ARPES with other ARPES variants are also underway. Two ARPES variants that have demonstrated great potential are spatially-resolved (micro or nano) ARPES and spin-resolved ARPES (SARPES). This section looks at the cross-over between these and TR-ARPES that have emerged in the literature.
\begin{figure}[t!]
    \centering
    \includegraphics{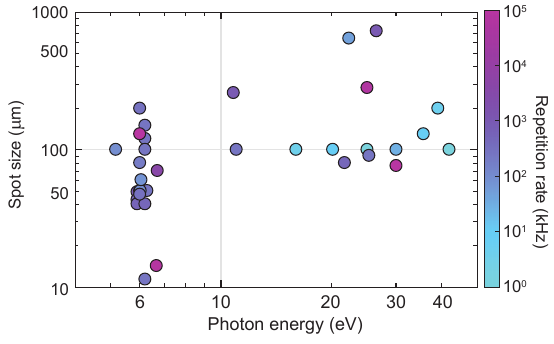}
    \caption{\textbf{Survey of probe sources spot sizes.} The optical spot size for various TR-ARPES sources. The 6-eV sources typically average smaller spot sizes from 30 to 200~$\upmu$m, while VUV/XUV sources typically have larger spot sizes on the order of 100~$\upmu$m as their lower repetition rates are more susceptible to space charge effects. The references are tabled in Tab.\,\ref{Tab: SS_survey}.}
    \label{Fig: SS}
\end{figure}

In Sec.\,\ref{Sec: space charge}, the ability of momentum-microscopes to perform TR-ARPES experiments with $\upmu$m spatial resolution was highlighted. In the absence of the momentum microscope, one can perform TR-$\upmu$ARPES by producing probes with spot size on the $\upmu$m scale. These small spot sizes are extremely useful for studying samples that are hard to grow uniformly over large areas or have multiple terminations upon cleaving. More crucially, they enable the study of two-dimensional exfoliated materials and heterostructures that can be challenging to make larger than $10-50~\upmu$m. Despite the short wavelength of probe sources, producing a small spot in-situ is not trivial. Complications arise due to long working distances for imaging into the UHV chamber, the quality of the probe beam, or the shape and quality of optics used to deliver the beam to the UHV system. Numerous examples exist of 6~eV sources producing spots of $<30~\upmu$m for high-resolution ARPES \cite{Kiss2008, Yan2021}, but many fewer reports of TR-ARPES exist \cite{Zhong2021, Dufresne2023}. Similarly, technical reports of focusing XUV to small spot sizes for equilibrium experiments exist, but relatively few time-resolved experiments were performed with spot sizes smaller than 100~$\upmu$m \cite{Corder2018, Maklar2022}. 

In Fig.\,\ref{Fig: SS}, a survey of probe spot size as a function of the photon energy is shown. The references are tabled in Tab.\,\ref{Tab: SS_survey}. We observe that the 6-eV sources typically operate from 30 to 200~$\upmu$m, while XUV sources operate with a larger spot size from 60 to 600~$\upmu$m. As discussed previously, one must balance the benefits of small spot size with the cost of increased space charge. At low repetition rates, the increased space charge will likely limit applications of TR-$\upmu$ARPES. On the other hand, the reduced probe beam size allows for a smaller pump beam, which reduces the average pump power delivered to the sample for a given pump fluence. This is desirable in very high repetition rate systems when the thermal load of the pump is a limitation. In the future, focusing closer to the diffraction limit with XUV light for TR-nano-ARPES may become practical with a high rep rate system. Micrometer spot sizes have been achieved in the UV \cite{Iwasawa2017} and soft X-rays with the latter using Fresnel zone plates (\cite{Kipp2001}) and focusing capillaries \cite{Koch2018}). In these experiments, the pulse duration is kept long (on the order of 10~ps) to achieve a narrow bandwidth and is used for high-energy resolution equilibrium measurements. However, even in this long pulse regime, the small spot size of the probe can exacerbate the SC dilemma by an order of magnitude \cite{Hellmann2012, Graf2010, Rotenberg2014}, which makes the already precarious balancing of parameters in TR-ARPES more arduous still. 

Another avenue that has been explored is the addition of spin resolution. Spin-resolved ARPES (SARPES) has been used in the study of magnetic systems, Rashba-split states, topological insulators, Weyl semimetals, and other material systems in which spin-orbit coupling is important. These experiments are typically very time-intensive due to the low efficiency of spin detectors, which rely on either the spin-orbit interaction (e.g. Mott detector) or spin-exchange interaction (e.g. VLEED detector) to selectively scatter electrons of a particular spin \cite{Okuda2017, Lin2021}. Though much effort has been put towards the development of multi-channel spin-detectors \cite{Shi2015}, modern commercial systems are zero-dimensional and are coupled to hemispherical or time-of-flight spectrometers for angular and energy resolution. Development of SARPES began in the 1960s, and the addition of time-resolution occurred alongside the development of two-photon photoemission and THG and HHG sources in the 1990s \cite{Scholl1997}. Today, spin and time-resolved ARPES (STARPES) have been demonstrated with both 6-eV \cite{Cacho2015, Jozwiak2016, Sanchez2017} and HHG sources \cite{Plotzing2016, Nie2019, Fanciulli2020}. However, the latter operates with relatively low repetition rates (see Tab.\,\ref{Tab: STARPES sources}). Given that SARPES experiments are already highly time-intensive, these STARPES systems would likely benefit from the development of higher repetition rate sources in the 100~kHz to MHz regime. 

 \begin{table}[t!]
\begin{center}
\caption{Table of laser sources used in STARPES setups. HA: Hemispherical analyzer, TOF: Time-of-flight, HHG: high harmonic generation, FHG: fourth-harmonic generation. FERRUM detectors are based on very low-energy electron diffraction (VLEED) from an oxygen-passivated Fe film. The spin-TOF is an integrated detector combining exchange scattering polarimetry with TOF electron spectroscopy.}
\label{Tab: STARPES sources}
\begin{tabular}{ c | c c c c  }
 Platform & Rep. Rate & Probe & Detector  & Year \\ 
 \hline
 Ti:sapphire \cite{Cacho2015} & 250~kHz & FHG, BBO & TOF-Mott \cite{Cacho2009} & 2015 \\  
 Ti:sapphire \cite{Jozwiak2016}  & 3.6~MHz & FHG, BBO & spin-TOF \cite{Jozwiak2010} & 2016\\    
Ti:sapphire \cite{Sanchez2017} & 100~kHz & FHG, BBO & HA-Mott & 2017\\
 Ti:sapphire \cite{Plotzing2016} & 3~kHz & HHG & FERRUM \cite{Escher2011} & 2016\\
 Ti:sapphire \cite{Nie2019} & 1~kHz & HHG & HA-Mott & 2019\\
 Ti:sapphire \cite{Fanciulli2020} & 10~kHz & HHG & FERRUM \cite{Escher2011} & 2020\\
\end{tabular}
\end{center}
\end{table}
\subsection{Free-electron Lasers}
One light source we have yet to address is the free-electron laser (FEL). Unlike the probe sources discussed thus far –-which are table-top laser-based sources-- FELs are large-scale user facilities with many beamlines performing a variety of experiments. For a detailed discussion of FELs, their technical capabilities and scientific contributions, we refer the reader to some excellent reviews \cite{Bostedt2016, Seddon2017, Feng2018, Rossbach2019}. The gain media in a FEL is relativistic electrons accelerating through an undulator. Here, both spontaneous and stimulated emission occurs, and the gain bandwidth is dependent on the electron energy, undulator period, and magnetic field strength of the undulator. FELs are capable of a huge range of photon energies from THz to hard X-ray. They are characterized by their high spectral brightness, coherence, and femtosecond pulse duration. To perform TR-ARPES with a FEL, the pump pulses must be synchronized to the FEL radiation, and this has been demonstrated with a timing jitter of less than 30~fs rms at FLASH using an all-optical locking technique \cite{Schulz2015}. To date, TR-ARPES studies have been performed in the soft X-ray regime at the FEL FLASH at DESY \cite{Kutnyakhov2020}, and in the hard X-ray regime at SACLA in RIKEN, Japan \cite{Oloff2014}. Efforts towards TR-ARPES are underway at other facilities, such as LCLS II and EuropeanXFEL. A detailed discussion of the design and light generation in an FEL is beyond the scope of this review; however, the parameters of the pump and probe parameters face the same fundamental and technical considerations of laboratory-based sources for TR-ARPES. 

\begin{figure}[t!]
    \centering
    \includegraphics{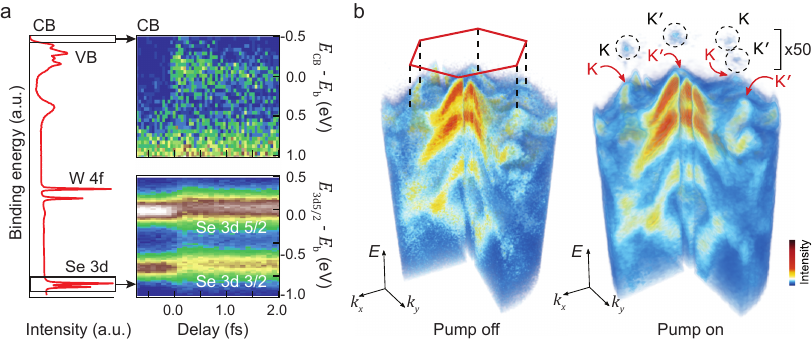}
    \caption{\textbf{An example of valence and core-level TR-ARPES at FLASH.} Figure adapted from Kutnyakhov \textit{et al.} \cite{Kutnyakhov2020}. \textbf{(a)} Left: Valence and core-level TR-ARPES at FLASH. Momentum-integrated spectra measured using 109.5~eV FEL probe pulses and s-polarized 775~nm pump pulses. Concurrent dynamics of the conduction band and core levels (Se $3d$ doublet) are shown on the right. \textbf{(b)} Volumetric rendering of the WSe$_2$ valence band before (left) and after (right) excitation with 775~nm pump pulses. The excited electrons at K and K' are clearly observed. The probe photon energy here is 36.5~eV.}
    \label{Fig: FLASH}
\end{figure}
In this respect, FELs can generate photons in the XUV to X-ray regime for valence and core-level TR-ARPES, respectively \cite{Kutnyakhov2020, Oloff2016, Oloff2014, Hellmann2012b, Pietzsch2008}. Unlike HHG sources, which are tunable in steps of twice the driving frequency, FEL radiation is continuously tunable over an extensive range in the soft and hard X-ray regime. This makes FELs the perfect tool for accessing electron dynamics at various points in the out-of-plane dispersion. The high photon energy of FELs also give the user the ability to study the ultrafast dynamics of core levels, whose spectral function can be sensitive to the formation of photoexcited hot carriers \cite{Curcio2021} and excitons \cite{Dendzik2020}. With respect to polarization tunability, the polarization control of XUV pulses generated at FLASH has been demonstrated \cite{vkSchmising2017}, though to our knowledge this capability is not yet integrated with TR-ARPES measurements. Lastly, the bandwidth and pulse duration of the probe pulse can be reasonably balanced, with Kutnyakhov \textit{et al.} \cite{Kutnyakhov2020} demonstrating values of 40~meV and 60~fs, respectively. 

In Fig.\,\ref{Fig: FLASH}, an example of the valence and core-level spectroscopy capability at FLASH is shown. Probe pulses with a photon energy of 109.5~eV are used to simultaneously measure the dynamics of valence and core levels after excitation (Fig.\,\ref{Fig: FLASH}a). In Fig.\,\ref{Fig: FLASH}b, 36.5~eV pulses are used to perform higher energy resolution studies of the valence band in WSe$_2$ and determine the excited state dynamics. Here, an MM-ARTOF generates a three-dimensional volumetric rendering of the valence band. After pump excitation with a 775~nm (1.6~eV) pulse, a clear population of electrons in the conduction band can be observed. Currently, FELs struggle somewhat with low repetition rates, with macro-bunches of 10 Hz and single-pulse repetition rates as high as 5 kHz. Consequently, TR-ARPES experiments at FELs generally contend with some space charge, limiting the ultimate energy resolution \cite{Hellmann2012b, Verna2016, Schonhense2018}. However, as efforts to increase the repetition rate are underway, this will surely improve.

\section{Concluding remarks}
In this manuscript, we have reviewed the development of laser sources for TR-ARPES and examined the intricate interdependencies of pump and probe source parameters with fundamental and technical aspects of TR-ARPES. In so doing, we have discussed necessary trade-offs in source parameters --such as pulse duration and bandwidth, pump fluence and repetition rate-- as well as the importance of the tunability of these parameters. These trade-offs are clearly observed in the survey of sources employed in TR-ARPES for both technical and scientific works. The survey figures (Fig.\,\ref{Fig: Probe_survey}, Fig.\,\ref{Fig: Probe_param}, Fig.\,\ref{Fig: Pump survey}, Fig.\,\ref{Fig: Pump param}) serve to identify some limitations of TR-ARPES sources, as well as regions of parameter space that have yet to be explored extensively. At the same time, the survey highlights the potential of TR-ARPES to access large regions of parameter space that are extremely relevant to many important scientific questions in condensed matter physics. 

What is also clear from these surveys is that the accessible parameter space directly depends on the details of pump and probe generation --from the laser platform through the subsequent stages of the nonlinear frequency conversion. For those seeking to build a TR-ARPES system, we have reviewed the prevalent options for pump and probe generation and their approach to the parameter trade-offs. Lastly, we discuss some future directions TR-ARPES based on the integration of source parameters that are not yet easily or widely tunable --such as pulse duration and probe polarization-- but which can be used to elucidate new non-equilibrium and equilibrium observables. The integration of pump-probe spectroscopy with ARPES variants such as $\upmu$ARPES and spin-resolved ARPES and the source parameters relevant to the increased complexity are also discussed. 

The proliferation of TR-ARPES in recent years is driven by its proven ability to access important scientific quantities and contribute insight to relevant scientific questions. In the next years, source development will grant even greater access to parameter space, and variables such as spin and orbital degrees of freedom will become readily available. While exciting, the increasing dimensionality of data presents fresh challenges in data acquisition, processing, and analysis, which are beginning to be addressed with a variety of machine learning techniques \cite{Peng2020, Kim2021, Melton2020, Iwasawa2022, Mortensen2023, Xian2023, Ekahana2023}. This review article --which focuses on the technical development of laser sources-- aims to highlight the collaboration of the ultrafast laser community and the condensed matter community that has made this possible.

\section{Acknowledgements}
We gratefully acknowledge our colleagues at the Quantum Matter Institute for their critical reading of the manuscript and the many thoughtful discussions throughout the years. Our thanks goes out to A. Damascelli, R. Day, J. Didap, S. Dufresne, B. Guislain, M. Hemsworth, C. Hofer, H.-H. Kung, G. Levy, I. Markovi\'{c}, M. Michiardi, M. Schneider, P. Sulzer, S. Smit, S. Zhdanovich, and B. Zwartsenberg. We also extend our thanks to members of the community for their critical reading and feedback on the manuscript: F. Boschini, F. Cilento, K. Dani, M. Dendzik, P. Hofmann, A. F. Kemper, P. Kirchmann, O. Tjernberg, S. Zhou, and M. Zonno. This research was undertaken thanks in part to funding from the Max PlanckUBC-UTokyo Centre for Quantum Materials and the Canada First Research Excellence Fund, Quantum Materials and Future Technologies Program. The work was supported by the Gordon and Betty Moore Foundation’s EPiQS Initiative, Grant GBMF4779, Canada Foundation for Innovation (CFI), British Columbia Knowledge Development Fund (BCKDF), and the Natural Sciences and Engineering Research Council of Canada (NSERC). 

\pagebreak
\appendix
\section{Tabulated references}
\label{Sec: Tables}
\fontsize{9}{11}\selectfont{
\begin{longtable}{ c|cccc }
\caption{The historical development of TR-ARPES as shown in Fig.\,\ref{Fig: history}. Post 2005, the proliferation of sources is such that this is a representative but not a complete list, especially for nonlinear-crystal based ``6-eV" sources. NLX: Nonlinear-crystal based sources. THG: Third harmonic generation in gases, driven by the 2nd or 3rd harmonic of the laser output. HHG: high harmonic generation in gases. 2/3H-HHG: HHG driven by the 2nd or 3rd harmonic of the laser output. Cavity-HHG: intracavity HHG. TOF: Time-of-flight. MA-TOF: Multi-anode TOF. MB-TOF: Magnetic bottle TOF. ARTOF: Angle-resolved TOF. MM-ARTOF: Momentum microscope-ARTOF. HA: Hemispherical analyzer.} \label{Tab: history}\\
\hline
First Author & Platform & Probe & Detector & Year \\
\hline
Haight \cite{Haight1985} & Dye & THG & TOF & 1985 \\
Haight \cite{Haight1988} & Dye & THG & MA-TOF & 1988 \\
Haight \cite{Haight1989} & Dye & THG & MA-TOF & 1989 \\
Fann \cite{Fann1992} & Dye & NLX & TOF & 1992 \\
Haight \cite{Haight1994} & Dye & HHG & MA-TOF & 1994 \\
Haight \cite{Haight1996} & Dye & HHG & MA-TOF & 1996 \\
Glover \cite{Glover1996} & Ti:sapphire & HHG & TOF & 1996 \\
Bouhal \cite{Bouhal1997} & Ti:sapphire & HHG & MB-TOF & 1997 \\
Scholl \cite{Scholl1997} & Ti:sapphire & NLX & Mott-detector & 1997 \\
Bouhal \cite{Bouhal1998} & Ti:sapphire & HHG & MB-TOF & 1998 \\
Link \cite{Link2001} & Ti:sapphire & NLX & HA & 2001 \\
Siffalovic \cite{Siffalovic2001} & Ti:sapphire & HHG & TOF & 2001 \\
Nugent-Glandorf \cite{Nugent-Glandorf2002} & Ti:sapphire & HHG & MB-TOF & 2002 \\
Bauer \cite{Bauer2003} & Ti:sapphire & HHG & TOF & 2003 \\
Rhie \cite{Rhie2003} & Ti:sapphire & NLX & TOF & 2003 \\
Lisowski \cite{Lisowski2005} & Ti:sapphire & NLX & TOF & 2005 \\
Miaja-Avila \cite{Miaja-Avila2006} & Ti:sapphire & HHG & TOF & 2006 \\
Mathias \cite{Mathias2007} & Ti:sapphire & HHG & HA & 2007 \\
Perfetti \cite{Perfetti2007} & Ti:sapphire & NLX & HA & 2007 \\
Perfetti \cite{Perfetti2008} & Ti:sapphire & NLX & ARTOF & 2008 \\
Schmitt \cite{Schmitt2008} & Ti:sapphire & NLX & ARTOF & 2008 \\
Dakovski \cite{Dakovski2010} & Ti:sapphire & HHG & HA & 2010 \\
Graf \cite{Graf2010} & Ti:sapphire & NLX & HA & 2010 \\
Frassetto \cite{Frassetto2011} & Ti:sapphire & HHG & HA & 2011 \\
Faure \cite{Faure2012} & Ti:sapphire & NLX & HA & 2012 \\
Sobota \cite{Sobota2012} & Ti:sapphire & NLX & HA & 2012 \\
Smallwood \cite{smallwood2012RSI} & Ti:sapphire & NLX & HA & 2012 \\
Crepaldi \cite{Crepaldi2012} & Ti:sapphire & NLX & HA & 2012 \\
Wang \cite{Wang2012} & Ti:sapphire & NLX & ARTOF & 2012 \\
Frietsch \cite{Frietsch2013} & Ti:sapphire & HHG & HA & 2013 \\
Eich \cite{Eich2014} & Ti:sapphire & 2H-HHG & HA & 2014 \\
Boschini \cite{Boschini2014} & Ytterbium & NLX & TOF & 2014 \\
Kanasaki \cite{Kanasaki2014} & Ti:sapphire & NLX & HA & 2014 \\
Ishida \cite{Ishida2014} & Ti:sapphire & NLX & HA & 2014 \\
Wegkamp \cite{Wegkamp2014} & Ti:sapphire & NLX & HA & 2014 \\
Ojeda \cite{Ojeda2016} & Ti:sapphire & HHG & TOF & 2015 \\
Dakovski \cite{Dakovski2015} & Ti:sapphire & HHG & HA & 2015 \\
Rohde \cite{Rohde2016} & Ti:sapphire & HHG & HA & 2016 \\
Cilento \cite{Cilento2016} & Ti:sapphire & THG & HA & 2016 \\
Ishida \cite{Ishida2016} & Ytterbium & NLX & HA & 2016 \\
S{\'a}nchez-Barriga \cite{Sanchez2016} & Ti:sapphire & NLX & HA & 2016 \\
Rettig \cite{Rettig2016} & Ti:sapphire & NLX & ARTOF & 2016 \\
Parham \cite{Parham2017} & Ti:sapphire & NLX & HA & 2017 \\
Boschini \cite{Boschini2018} & Ti:sapphire & NLX & HA & 2018 \\
Reimann \cite{Reimann2018} & Ti:sapphire & NLX & HA & 2018 \\
Corder \cite{Corder2018} & Ytterbium & cavity-HHG & HA & 2018\\
Buss \cite{Buss2019} & Ti:sapphire & 2H-HHG & HA & 2019 \\
Sie \cite{Sie2019} & Ti:sapphire & HHG & ARTOF & 2019 \\
Puppin \cite{Puppin2019} & Ytterbium & 2H-HHG & HA & 2019 \\
Freutel \cite{Freutel2019} & Ti:sapphire & NLX & ARTOF & 2019 \\
Roth \cite{Roth2019} & Ti:sapphire & HHG & HA & 2019 \\
Saule \cite{Saule2019} & Ytterbium & cavity-HHG & ARTOF & 2019 \\
Yang \cite{Yang2019} & Ytterbium & NLX & HA & 2019 \\
Mills \cite{Mills2019} & Ytterbium & cavity-HHG & HA & 2019 \\
Peli \cite{Peli2020} & Ytterbium & THG & HA & 2020 \\
Lee \cite{Lee2020} & Ytterbium & THG & ARTOF & 2020 \\
Keunecke \cite{Keunecke2020} & Ytterbium & 2H-HHG & MM-ARTOF & 2020 \\
Gauthier \cite{Gauthier2020} & Ti:sapphire & NLX & HA & 2020 \\
Mad{\'e}o \cite{Madeo2020} & Ytterbium & 2H-HHG & MM-ARTOF & 2020 \\
Cucini \cite{Cucini2020} & Ytterbium & 2H-HHG & HA & 2020 \\
Bao \cite{Bao2021} & Ti:sapphire & NLX & HA & 2021 \\
Yan \cite{Yan2021} & Ytterbium & NLX & HA & 2021 \\
Bao \cite{Bao2022} & Ti:sapphire & NLX, KBBF & HA & 2022 \\
Guo \cite{Guo2022} & Ytterbium & 3H-HHG & ARTOF & 2022 \\
Heber \cite{Heber2022} & Ti:sapphire & HHG & MM-ARTOF & 2022 \\
Zhong \cite{Zhong2022} & Ti:sapphire & NLX, KBBF & HA & 2022 \\
Nevola \cite{Nevola2023} & Ti:sapphire & NLX & HA & 2022 \\
Kunin \cite{Kunin2023} & Ytterbium & cavity-HHG & MM-ARTOF & 2022 \\
Maklar \cite{Maklar2022} & Ytterbium & 3H-HHG & MM-ARTOF & 2022 \\
D\"{u}vel \cite{Duvel2022} & Ytterbium & 2H-HHG & HA & 2022 \\
Kawaguchi \cite{Kawaguchi2023} & Ytterbium & THG & HA & 2023
\end{longtable}
}
\pagebreak
\fontsize{8}{10}\selectfont{
\begin{longtable}{ c|cccccc }
\caption{Survey of the probe source parameters plotted in Fig.\,\ref{Fig: Probe_survey} and \ref{Fig: Probe_param}. ER: Energy resolution, typically obtained from a fit of the Fermi-edge on a cold polycrystalline gold. TR: Time resolution, defined as the cross-correlation of the pump and the probe pulse. Sources with demonstrated tunability in the photon energy or repetition rate are indicated with an asterisk. Ti:sapphire sources are listed first, followed by ytterbium sources. In some cases, we have estimated the time resolution from the reported pump and probe pulse duration.} \label{Tab: probe_survey}\\
First Author & Platform & Probe $h\nu$ (eV) & ER (meV) & TR (fs) & Rep. Rate & Year \\
\hline
Frassetto   \cite{Frassetto2011} & Ti:sapphire & 32.5* & 700 & 30 & 1 kHz & 2011 \\
Petersen \cite{Petersen2011} & Ti:sapphire & 20.4* & 150 & 30 & 1 kHz & 2011 \\
Gierz \cite{Gierz2013} & Ti:sapphire & 31.5* & 130 & 50 & 1 kHz & 2013 \\
Ulstrup \cite{Ulstrup2015} & Ti:sapphire & 33.2* & 350 & 60 & 1 kHz & 2015 \\
Ulstrup \cite{Ulstrup2016} & Ti:sapphire & 25* & 400 & 40 & 1 kHz & 2016 \\
Crepaldi \cite{Crepaldi2017} & Ti:sapphire & 17.5* & 150 & 50 & 1 kHz & 2017 \\
Smallwood \cite{smallwood2012RSI} & Ti:sapphire & 5.9 & 23 & 310 & 543 kHz* & 2012 \\
Buss \cite{Buss2019} & Ti:sapphire & 22.3 & 60.4 & 65 & 50 kHz* & 2019 \\
Parham \cite{Parham2017} & Ti:sapphire & 6.3 & 9 & 700 & 20 kHz & 2017 \\
Freutel \cite{Freutel2019} & Ti:sapphire & 6 & 55 & 100 & 250 kHz & 2019 \\
Ojeda \cite{Ojeda2016} & Ti:sapphire & 39.2* & 200 & 180 & 6 kHz* & 2016 \\
Crepaldi \cite{Crepaldi2022} & Ti:sapphire & 27* & 200 & 60 & 6 kHz* & 2022 \\
Roth \cite{Roth2019} & Ti:sapphire & 37* & 130 & 200 & 6 kHz* & 2019 \\
Faure \cite{Faure2012} & Ti:sapphire & 6.28 & 70 & 65 & 250 kHz & 2012 \\
Zhang \cite{Zhang2021} & Ti:sapphire & 6.2 & 15 & 160 & 500 kHz & 2021 \\
Perfetti \cite{Perfetti2007} & Ti:sapphire & 6 & 47 & 100 & 30 kHz & 2007 \\
Frietsch \cite{Frietsch2013} & Ti:sapphire & 35.6* & 90 & 125 & 10 kHz & 2013 \\
Rameau \cite{Rameau2016} & Ti:sapphire & 6.2 & 55 & 100 & 250 kHz & 2016 \\
Frietsch \cite{Frietsch2020} & Ti:sapphire & 36.8* & 200 & 120 & 10 kHz & 2020 \\
Monney \cite{Monney2018} & Ti:sapphire & 6 & 50 & 250 & 30 kHz & 2018 \\
Wegkamp \cite{Wegkamp2014} & Ti:sapphire & 6.19 & 90 & 89 & 40 kHz & 2014 \\
S{\'a}nchez-Barriga \cite{Sanchez2016} & Ti:sapphire & 6 & 30 & 200 & 150 kHz & 2016 \\
Mathias \cite{Mathias2007} & Ti:sapphire & 41.85* & 800 & 20 & 1 kHz & 2007 \\
Eich \cite{Eich2014} & Ti:sapphire & 22.3 & 150 & 30 & 10 kHz* & 2014 \\
Stange \cite{Stange2015} & Ti:sapphire & 22.1 & 240 & 32 & 8 kHz* & 2015 \\
Zhang \cite{Zhang2020} & Ti:sapphire & 22.4 & 130 & 41 & 4 kHz & 2020 \\
Emmerich \cite{Emmerich2020} & Ti:sapphire & 22.2 & 150 & 30 & 10 kHz & 2020 \\
Pl{\"o}tzing  \cite{Plotzing2016} & Ti:sapphire & 42.7* & 1000 & 35 & 3 kHz & 2016 \\
Pl{\"o}tzing  \cite{Plotzing2016} & Ti:sapphire & 22.5* & 350 & 35 & 5 kHz & 2016 \\
Rohwer \cite{Rohwer2011} & Ti:sapphire & 43* & 400 & 33 & 3 kHz & 2011 \\
Rohde \cite{Rohde2016} & Ti:sapphire & 22.1 & 170 & 32 & 10 kHz & 2016 \\
Dakovski \cite{Dakovski2015} & Ti:sapphire & 20.15* & 250 & 35 & 10 kHz & 2015 \\
Wallauer \cite{Wallauer2021} & Ti:sapphire & 21.7* & 100 & 50 & 200 kHz & 2021 \\
Sie \cite{Sie2019} & Ti:sapphire & 30* & 30 & 200 & 30 kHz & 2019 \\
Kanasaki \cite{Kanasaki2014} & Ti:sapphire & 4.5 & 50 & 98 & 76 MHz & 2014 \\
Fanciulli \cite{Fanciulli2020} & Ti:sapphire & 35.65* & 250 & 30 & 10 kHz & 2020 \\
Reimann \cite{Reimann2018} & Ti:sapphire & 6.2 & 45 & 100 & 3 kHz & 2018 \\
Aeschlimann \cite{Aeschlimann2021} & Ti:sapphire & 21.7 & 150 & 300 & 1 kHz & 2021 \\
Sobota \cite{Sobota2012} & Ti:sapphire & 6 & 22 & 163 & 80 MHz & 2012 \\
Gauthier \cite{Gauthier2020} & Ti:sapphire & 6 & 55-27 & 58-103 & 312 kHz & 2020 \\
Crepaldi \cite{Crepaldi2012} & Ti:sapphire & 6.2 & 10 & 300 & 250 kHz & 2012 \\
Bao \cite{Bao2021} & Ti:sapphire & 5.9 & 58 & 81 & 76 MHz & 2021 \\
Zhong \cite{Zhong2022} & Ti:sapphire & 5.8* & 67 & 81 & 76 MHz* & 2022 \\
Zhong \cite{Zhong2022} & Ti:sapphire & 6.2* & 35 & 86 & 76 MHz* & 2022 \\
Zhong \cite{Zhong2022} & Ti:sapphire & 6.7* & 43 & 88 & 76 MHz* & 2022 \\
Zhong \cite{Zhong2022} & Ti:sapphire & 6.9* & 62 & 95 & 76 MHz* & 2022 \\
Bao \cite{Bao2022b} & Ti:sapphire & 6.7* & 16 & 480 & 3.8 MHz* & 2022 \\
Boschini \cite{Boschini2018} & Ti:sapphire & 6.2 & 19 & 250 & 250 kHz & 2018 \\
Dufresne \cite{Dufresne2023} & Ti:sapphire & 6.2 & 11 & 280 & 250 kHz & 2023 \\
Ishida \cite{Ishida2014} & Ti:sapphire & 5.92 & 10.5 & 240 & 250 kHz* & 2014 \\
Mitsuoka \cite{Mitsuoka2020} & Ti:sapphire & 21.7 & 250 & 80 & 10 kHz & 2020 \\
Suzuki \cite{Suzuki2021} & Ti:sapphire & 21.7 & 250 & 70 & 10 kHz & 2021 \\
Watanabe \cite{Watanabe2022} & Ti:sapphire & 27.9* & 250 & 80 & 1 kHz & 2022 \\
Heber \cite{Heber2022} & Ti:sapphire & 36.1* & 96 & 95 & 6 kHz & 2022 \\
Plogmaker \cite{Plogmaker2015} & Ti:sapphire & 39* & 110 & 60 & 5 kHz* & 2015 \\
Kremer \cite{Kremer2021} & Ytterbium & 6.3 & 47 & 100 & 200 kHz & 2021 \\
Puppin \cite{Puppin2019} & Ytterbium & 21.7 & 121 & 37 & 500 kHz & 2019 \\
Maklar \cite{Maklar2022} & Ytterbium & 21.7 & 150 & 41 & 500 kHz & 2022 \\
Keunecke \cite{Keunecke2020} & Ytterbium & 26.5 & 200 & 42 & 1 MHz & 2020 \\
D{\"u}vel \cite{Duvel2022} & Ytterbium & 21.7 & 220 & 45 & 500 kHz & 2022 \\
Hein \cite{Hein2020} & Ytterbium & 5.9 & 40 & 100 & 500 kHz & 2020 \\
Lee \cite{Lee2020} & Ytterbium & 11 & 16 & 250 & 250 kHz* & 2020 \\
Mad{\'e}o \cite{Madeo2020} & Ytterbium & 21.7 & 30 & 240 & 1 MHz* & 2020 \\
Boschini \cite{Boschini2014} & Ytterbium & 6.02 & 50 & 85 & 100 kHz* & 2014 \\
Bugini \cite{Bugini2017} & Ytterbium & 6.05 & 80 & 72 & 80 kHz* & 2017 \\
Yang \cite{Yang2019} & Ytterbium & 6.05 & 19 & 130 & 500 kHz* & 2019 \\
Corder \cite{Corder2018} & Ytterbium & 30* & 110 & 181 & 88 MHz* & 2018 \\
Peli \cite{Peli2020} & Ytterbium & 10.8 & 26 & 700 & 1 MHz* & 2020 \\
Mills \cite{Mills2019} & Ytterbium & 25* & 22 & 190 & 60 MHz & 2019 \\
Yan \cite{Yan2021} & Ytterbium & 6 & 17 & 115 & 200 kHz & 2021 \\
Ishida \cite{Ishida2016} & Ytterbium & 6 & 11.3 & 310 & 95 MHz & 2016 \\
Guo \cite{Guo2022} & Ytterbium & 18.1* & 14 & 204 & 250 kHz & 2022 \\
Guo \cite{Guo2022} & Ytterbium & 25.3* & 18 & 165 & 250 kHz & 2022
\end{longtable}
}
\pagebreak
\begin{longtable}{ c|ccccc }
\caption{Survey of the pump source parameters shown in Fig.\,\ref{Fig: Pump survey} and \ref{Fig: Pump param}. Absorbed fluence values are indicated with an asterisk, while incident fluence values are without an asterisk. In some cases, we have estimated the fluence from the reported pulse energy and spot size. Sources using a UV (6-eV) probe are listed first, followed by those using an XUV probe.} \label{Tab: pump_survey}\\
First Author & Platform & Pump $h\nu$ (eV) & Fluence~($\upmu$J/cm$^{2}$) & Rep. Rate & Year \\
\hline
Graf   \cite{Graf2011} & Ti:sapphire & 1.48 & 40 & 905 kHz & 2011 \\
Smallwood   \cite{Smallwood2012} & Ti:sapphire & 1.48,  1.48 & 2,  15 & 543 kHz & 2012 \\
Parham \cite{Parham2017} & Ti:sapphire & 0.31,  0.31,  0.31 & 45,  60,  80 & 20 kHz & 2017 \\
Yang \cite{Yang2017} & Ti:sapphire & 1.57 & 130 & 20 kHz & 2017 \\
Freutel \cite{Freutel2019} & Ti:sapphire & 1.5 & 200 & 250 kHz & 2019 \\
Nevola \cite{Nevola2023} & Ti:sapphire & 1.55,  1.55,  1.55,    1.55 & 1,  2,  3,  4 & 250 kHz & 2022 \\
Perfetti \cite{Perfetti2008} & Ti:sapphire & 1.5 & 135* & 200 kHz & 2008 \\
Hajlaoui \cite{Hajlaoui2014} & Ti:sapphire & 1.57 & 100 & 250 kHz & 2014 \\
Nilforoushan   \cite{Nilforoushan2020} & Ti:sapphire & 1.55,  1.55,  1.55 & 200,  270,  500 & 250 kHz & 2020 \\
Zhang \cite{Zhang2021} & Ti:sapphire & 1.55 & 120 & 500 kHz & 2021 \\
Perfetti \cite{Perfetti2007} & Ti:sapphire & 1.5 & 100 & 30 kHz & 2007 \\
Schmitt \cite{Schmitt2008} & Ti:sapphire & 1.5,  1.5 & 300,  2000 & 300 kHz & 2008 \\
Cort{\'e}s \cite{Cortes2011} & Ti:sapphire & 1.5,  1.5,  1.5,    1.5 & 6,  13,  32,    139* & 300 kHz & 2011 \\
Schmitt \cite{Schmitt2011} & Ti:sapphire & 1.5,  1.5 & 50,  150 & 300 kHz & 2011 \\
Avigo \cite{Avigo2017} & Ti:sapphire & 1.5,  1.5 & 800,  1400 & 300 kHz & 2017 \\
Rettig \cite{Rettig2013} & Ti:sapphire & 1.5 & 800 & 300 kHz & 2013 \\
Rameau \cite{Rameau2016} & Ti:sapphire & 1.5, 1.5, 1.5 & 35, 105, 315* & 250 kHz & 2016 \\
Monney \cite{Monney2018} & Ti:sapphire & 1.5,  1.5 & 1900,  2300* & 30 kHz & 2018 \\
Wegkamp \cite{Wegkamp2014} & Ti:sapphire & 1.54 & 6700 & 40 kHz & 2014 \\
Mor \cite{Mor2017} & Ti:sapphire & 1.55,  1.55 & 30,  120 & 40 kHz & 2017 \\
S{\'a}nchez-Barriga \cite{Sanchez2017} & Ti:sapphire & 1.5 & 100 & 150 kHz & 2017 \\
Varykhalov   \cite{Varykhalov2020} & Ti:sapphire & 1.5 & 100 & 150 kHz & 2020 \\
Hein \cite{Hein2020} & Ytterbium & 1.5 & 110 & 500 kHz & 2020 \\
Kuroda \cite{Kuroda2016} & Ti:sapphire & 0.25,  0.37 & 100,  100 & 100 kHz & 2016 \\
Wang \cite{Wang2013} & Ti:sapphire & 0.12 & 300 & 30 kHz & 2013 \\
Mahmood \cite{Mahmood2016} & Ti:sapphire & 0.16 & 600 & 30 kHz & 2016 \\
Baldini \cite{Baldini2023} & Ti:sapphire & 1.55,  1.55 & 500,  850* & 30 kHz & 2020 \\
Bugini \cite{Bugini2017} & Ytterbium & 1.85 & 500 & 80 kHz & 2017 \\
Hedayat \cite{Hedayat2019} & Ytterbium & 1.82,  1.82, 1.82, 1.82 & 31,  62, 125, 250 & 80 kHz & 2019 \\
Sayers \cite{Sayers2020} & Ytterbium & 1.8 & 1100 & 80 kHz & 2020 \\
Hedayat \cite{Hedayat2021} & Ytterbium & 1.82,  1.82 & 24,  400 & 80 kHz & 2021 \\
Tang \cite{Tang2020} & Ytterbium & 1.77,  1.77 & 50,  320 & 500 kHz & 2020 \\
Yang \cite{Yang2014} & Ti:sapphire & 1.5 & 25 & 80 MHz & 2014 \\
Gerber \cite{Gerber2017} & Ti:sapphire & 1.5,  1.5 & 120,  620 & 312 kHz & 2017 \\
Yang \cite{Yang2015} & Ti:sapphire & 1.5,  1.5,  1.5 & 10,  27,  60 & 800 kHz & 2015 \\
Crepaldi \cite{Crepaldi2012} & Ti:sapphire & 1.55 & 210* & 250 kHz & 2012 \\
Manzoni \cite{Manzoni2015} & Ti:sapphire & 1.55 & 100 & 250 kHz & 2015 \\
Zhong \cite{Zhong2021} & Ti:sapphire & 1.66,  1.66 & 64,  224 & 76 MHz & 2021 \\
Bao \cite{Bao2022} & Ti:sapphire & 1.68 & 40 & 3.8 MHz & 2022 \\
Zhou \cite{Zhou2023} & Ti:sapphire & 0.25, 0.34, 0.43, 0.5 & 700, 700, 700, 700 & 10 kHz & 2022 \\
Boschini \cite{Boschini2018} & Ti:sapphire & 1.55,  1.55 & 8,  30 & 250 kHz & 2018 \\
Boschini \cite{Boschini2020} & Ti:sapphire & 1.55,  1.55 & 28,  50 & 250 kHz & 2020 \\
Michiardi   \cite{Michiardi2022} & Ti:sapphire & 1.55,  1.55 & 40,  80 & 250 kHz & 2022 \\
Gole{\v z} \cite{Golez2022} & Ti:sapphire & 1.55,  1.55 & 26,  160 & 250 kHz & 2022 \\
Yan \cite{Yan2021} & Ytterbium & 1.55 & 200 & 200 kHz & 2021 \\
Ishida \cite{Ishida2016} & Ytterbium & 1.2 & 3 & 95 MHz & 2016 \\
Petersen \cite{Petersen2011} & Ti:sapphire & 1.57 & 500 & 1 kHz & 2011 \\
Liu \cite{Liu2013} & Ti:sapphire & 1.57 & 500 & 1 kHz & 2013 \\
Gierz \cite{Gierz2013} & Ti:sapphire & 0.3, 0.95 & 800, 4600 & 1 kHz & 2013 \\
Gierz \cite{Gierz2015} & Ti:sapphire & 0.3, 0.2, 0.17, 0.137 & 260, 260, 260, 260 & 1 kHz & 2015 \\
Ulstrup \cite{Ulstrup2015} & Ti:sapphire & 0.95, 1.6 & 340, 31000 & 1 kHz & 2015 \\
Ulstrup \cite{Ulstrup2016} & Ti:sapphire & 1.6, 2.0, 3.0 & 700, 3000, 1300 & 1 kHz & 2016 \\
Ulstrup \cite{Ulstrup2017} & Ti:sapphire & 2.05 & 3000 & 1 kHz & 2017 \\
Crepaldi \cite{Crepaldi2017} & Ti:sapphire & 2 & 300 & 1 kHz & 2017 \\
Cilento \cite{Cilento2018} & Ti:sapphire & 0.82, 1.65 & 350, 350 & 1 kHz & 2018 \\
Andreatta   \cite{Andreatta2019} & Ti:sapphire & 2.05, 2.05, 2.05 & 4400, 7800, 8400 & 1 kHz & 2019 \\
Lee \cite{Lee2021} & Ti:sapphire & 2.2 & 40 & 25 kHz & 2021 \\
Huber \cite{Huber2022} & Ti:sapphire & 1.6 & 80 & 50 kHz & 2022 \\
Roth \cite{Roth2019} & Ti:sapphire & 1.55 & 150* & 6 kHz & 2019 \\
Gatti \cite{Gatti2020} & Ti:sapphire & 1.6 & 380* & 6 kHz & 2020 \\
Frietsch \cite{Frietsch2020} & Ti:sapphire & 1.6,  1.6 & 2500,  3500* & 10 kHz & 2020 \\
Beaulieu \cite{Beaulieu2021} & Ytterbium & 1.2 & 600* & 500 kHz & 2021 \\
Maklar \cite{Maklar2022} & Ytterbium & 1.55, 3.1 & 140, 500 & 500 kHz & 2022 \\
Puppin \cite{Puppin2022} & Ti:sapphire & 3.1, 3.1 & 10, 350 & 500 kHz & 2022 \\
Schmitt \cite{Schmitt2022} & Ytterbium & 1.7 & 280 & 1 MHz & 2022 \\
D{\" u}vel \cite{Duvel2022} & Ytterbium & 1.2 & 1300 & 500 kHz & 2022 \\
Stange \cite{Stange2015} & Ti:sapphire & 1.6, 1.6, 1.6 & 35, 280, 1400 & 8 kHz & 2015 \\
Tengdin \cite{Tengdin2018} & Ti:sapphire & 1.6, 1.6 & 500, 14000 & 4 kHz & 2018 \\
Zhang \cite{Zhang2020} & Ti:sapphire & 1.6, 1.6 & 240, 860 & 4 kHz & 2020 \\
Emmerich \cite{Emmerich2020} & Ti:sapphire & 3.2, 3.2 & 10, 100 & 10 kHz & 2020 \\
Rohwer \cite{Rohwer2011} & Ti:sapphire & 1.57, 1.57 & 200, 5000 & 3 kHz & 2011 \\
Dakovski \cite{Dakovski2015} & Ti:sapphire & 1.55 & 600* & 10 kHz & 2015 \\
Wallauer \cite{Wallauer2016} & Ti:sapphire & 2.05 & 1000 & 100 kHz & 2016 \\
Wallauer \cite{Wallauer2021} & Ti:sapphire & 2.03 & 50 & 200 kHz & 2021 \\
Sie \cite{Sie2019} & Ti:sapphire & 1.59 & 1200 & 30 kHz & 2019 \\
Lee \cite{Lee2020} & Ytterbium & 1.82, 1.72 & 400, 700 & 250 kHz & 2020 \\
Baldini \cite{Baldini2023} & Ytterbium & 1.55 & 400* & 100 kHz & 2020 \\
Mad{\'e}o \cite{Madeo2020} & Ytterbium & 1.72, 2.48 & 14, 5* & 2 MHz & 2020 \\
Aeschlimann   \cite{Aeschlimann2021} & Ti:sapphire & 0.28 & 2000 & 1 kHz & 2021 \\
Kunin \cite{Kunin2023} & Ytterbium & 2.4, 2.4 & 1.3, 29 & 61 MHz & 2022 \\
Na \cite{Na2019} & Ytterbium & 1.2 & 20 & 60 MHz & 2019 \\
Mitsuoka \cite{Mitsuoka2020} & Ti:sapphire & 1.55 & 950 & 10 kHz & 2020 \\
Suzuki \cite{Suzuki2021} & Ti:sapphire & 1.55 & 2270 & 10 kHz & 2021 \\
Watanabe \cite{Watanabe2022} & Ti:sapphire & 1.55 & 2000 & 1 kHz & 2022 
\end{longtable}
\pagebreak
\fontsize{9}{11}\selectfont{
\begin{longtable}{ c|cccc }
\caption{Survey of the probe spot sizes as shown in Fig.\,\ref{Fig: SS}. These values are usually the full-width-at-half-maximum (FWHM), though in some cases an ambiguous spot diameter is given. Where both dimensions of spot size are reported, the plotted value is the square root of the product. In most of these setups, the probe spot size is equal to or smaller than the pump spot size. In studies where the opposite is true, an MM-ARTOF is used for spatial filtering, and these are indicated with an asterisk.} \label{Tab: SS_survey}\\
First Author & Probe h$\nu$ (eV) & Spot size ($\upmu$m) & Rep. Rate & Year \\
\hline
Andreatta   \cite{Andreatta2019} & 25 & 100 & 1 kHz & 2019 \\
Graf \cite{Graf2011} & 5.9 & 40~$\times$  60 & 905 kHz & 2011 \\
Smallwood \cite{smallwood2012RSI} & 5.9 & 43 & 543 kHz & 2012 \\
Smallwood \cite{Smallwood2012} & 5.9 & 40 & 543 kHz & 2012 \\
Buss \cite{Buss2019} & 22.3 & 650 & 50 kHz & 2019 \\
Freutel \cite{Freutel2019} & 6 & 50 & 250 kHz & 2019 \\
Perfetti \cite{Perfetti2008} & 6 & 80 & 200 kHz & 2008 \\
Faure \cite{Faure2012} & 6.28 & 50 & 250 kHz & 2012 \\
Zhang \cite{Zhang2021} & 6.2 & 40 & 500 kHz & 2021 \\
Schmitt \cite{Schmitt2008} & 6 & 200 & 300 kHz & 2008 \\
Frietsch \cite{Frietsch2013} & 35.6 & 130 & 10 kHz & 2013 \\
Maklar \cite{Maklar2022} & 21.7 & 80~$\times$  80 & 500 kHz & 2022 \\
Keunecke \cite{Keunecke2020} & 26.5 & 600~$\times$  900* & 1 MHz & 2020 \\
Mathias \cite{Mathias2007} & 41.85 & 100 & 1 kHz & 2007 \\
Tengdin \cite{Tengdin2018} & 16 & 100 & 4 kHz & 2018 \\
Dakovski \cite{Dakovski2015} & 20.15 & 100 & 10 kHz & 2015 \\
Kuroda \cite{Kuroda2016} & 5.16 & 100 & 100 kHz & 2016 \\
Sie \cite{Sie2019} & 30 & 100 & 30 kHz & 2019 \\
Lee \cite{Lee2020} & 11 & 100 & 250 kHz & 2020 \\
Boschini \cite{Boschini2014} & 6.02 & 50 & 100 kHz & 2014 \\
Hedayat \cite{Hedayat2021} & 6.05 & 60 & 80 kHz & 2021 \\
Gerber \cite{Gerber2017} & 6 & 34~$\times$  65 & 312 kHz & 2017 \\
Corder \cite{Corder2018} & 30 & 58~$\times$  100 & 88 MHz & 2018 \\
Peli \cite{Peli2020} & 10.7 & 260 & 1 MHz & 2020 \\
Zhong \cite{Zhong2021} & 6.66 & 10~$\times$  20 & 76 MHz & 2021 \\
Bao \cite{Bao2022b} & 6.7 & 70 & 3.8 kHz & 2022 \\
Mills \cite{Mills2019} & 25 & 200~$\times$  400 & 60 MHz & 2019 \\
Boschini \cite{Boschini2018} & 6.2 & 120 & 250 kHz & 2018 \\
Boschini \cite{Boschini2020} & 6.2 & 150 & 250 kHz & 2020 \\
Dufresne \cite{Dufresne2023} & 6.2 & 10.5~$\times$ 12.1 & 250 kHz & 2023 \\
Michiardi \cite{Michiardi2022} & 6.2 & 100 & 250 kHz & 2022 \\
Ishida \cite{Ishida2016} & 6 & 130 & 95 MHz & 2016 \\
Guo \cite{Guo2022} & 25.3 & 96~$\times$  85 & 250 kHz & 2022 \\
Plogmaker \cite{Plogmaker2015} & 39 & 200 & 5 kHz & 2015 
\end{longtable}
}
\pagebreak






\end{document}